\documentclass[aps,prd,twocolumn,superscriptaddress,showpacs]{revtex4}
\usepackage{graphicx}
\usepackage{subfigure}

\newcommand{\Bot}{$b$}
\newcommand{\Charm}{$c$}
\newcommand{\ttbar}{$t\bar{t}$}
\newcommand{\bbbar}{$b\bar{b}$}
\newcommand{\ccbar}{$c\bar{c}$}

\newcommand{\pT}{$p_T$}
\newcommand{\W}{$W$}
\newcommand{\Z}{$Z$}
\newcommand{\ifb}{$\textnormal{fb}^{-1}$}
\newcommand{\ipb}{$\textnormal{pb}^{-1}$}
\newcommand{\lumi}{$1.9\ \textnormal{fb}^{-1}$}
\newcommand{\lxyCv}{166.9 }
\newcommand{\lxyUpStatErr}{9.5 }
\newcommand{\lxyDownStatErr}{8.5 }
\newcommand{\lxySystErr}{ 2.9 }
\newcommand{\lepPtCv}{173.5 }
\newcommand{\lepPtUpStatErr}{8.8}
\newcommand{\lepPtDownStatErr}{8.9}
\newcommand{\lepPtSystErr}{ 3.8 }
\newcommand{\combCv}{170.7 }
\newcommand{\combStatErr}{6.3 }
\newcommand{\combSystErr}{ 2.6 }
\newcommand{\lxyRes}{$ \lxyCv ^{+\lxyUpStatErr}_{-\lxyDownStatErr}\ \textnormal{(stat)} \pm \lxySystErr\ \textnormal{(syst)\ GeV} / c^2$}
\newcommand{\lepPtRes}{$ \lepPtCv ^{+\lepPtUpStatErr}_{- \lepPtDownStatErr}\ \textnormal{(stat)} \pm \lepPtSystErr\ \textnormal{(syst)\ GeV} / c^2 $}
\newcommand{\combRes}{$ \combCv \pm \combStatErr\ \textnormal{(stat)} \pm \combSystErr\ \textnormal{(syst)}\ \textnormal{GeV} / c^2$}
\newcommand{\lxyStatRes}{$ \lxyCv ^{+\lxyUpStatErr}_{-\lxyDownStatErr}\  \textnormal{GeV} / c^2$}
\newcommand{\lepPtStatRes}{$ \lepPtCv ^{+\lepPtUpStatErr}_{- \lepPtDownStatErr}\  \textnormal{GeV} / c^2 $}
\newcommand{\combStatRes}{ $ \combCv \pm \combStatErr \ \textnormal{GeV }/ c^2 $}
\newcommand{\dataLepPt}{$55.2 \pm 1.3\ \textnormal{GeV} / c$}
\newcommand{\dataLxy}{$0.590 \pm 0.017\ \textnormal{cm}$}

\newcommand{\lepPtSystLim}{$3.0\ \textnormal{GeV}/c^2$}
\newcommand{\combSystLim}{$2.3\ \textnormal{GeV}/c^2$}

\begin{document}

\title{Measurements of the top-quark mass using charged particle tracking}


\affiliation{Institute of Physics, Academia Sinica, Taipei, Taiwan 11529, Republic of China} 
\affiliation{Argonne National Laboratory, Argonne, Illinois 60439} 
\affiliation{University of Athens, 157 71 Athens, Greece} 
\affiliation{Institut de Fisica d'Altes Energies, Universitat Autonoma de Barcelona, E-08193, Bellaterra (Barcelona), Spain} 
\affiliation{Baylor University, Waco, Texas  76798} 
\affiliation{Istituto Nazionale di Fisica Nucleare Bologna, $^y$University of Bologna, I-40127 Bologna, Italy} 
\affiliation{Brandeis University, Waltham, Massachusetts 02254} 
\affiliation{University of California, Davis, Davis, California  95616} 
\affiliation{University of California, Los Angeles, Los Angeles, California  90024} 
\affiliation{University of California, San Diego, La Jolla, California  92093} 
\affiliation{University of California, Santa Barbara, Santa Barbara, California 93106} 
\affiliation{Instituto de Fisica de Cantabria, CSIC-University of Cantabria, 39005 Santander, Spain} 
\affiliation{Carnegie Mellon University, Pittsburgh, PA  15213} 
\affiliation{Enrico Fermi Institute, University of Chicago, Chicago, Illinois 60637}
\affiliation{Comenius University, 842 48 Bratislava, Slovakia; Institute of Experimental Physics, 040 01 Kosice, Slovakia} 
\affiliation{Joint Institute for Nuclear Research, RU-141980 Dubna, Russia} 
\affiliation{Duke University, Durham, North Carolina  27708} 
\affiliation{Fermi National Accelerator Laboratory, Batavia, Illinois 60510} 
\affiliation{University of Florida, Gainesville, Florida  32611} 
\affiliation{Laboratori Nazionali di Frascati, Istituto Nazionale di Fisica Nucleare, I-00044 Frascati, Italy} 
\affiliation{University of Geneva, CH-1211 Geneva 4, Switzerland} 
\affiliation{Glasgow University, Glasgow G12 8QQ, United Kingdom} 
\affiliation{Harvard University, Cambridge, Massachusetts 02138} 
\affiliation{Division of High Energy Physics, Department of Physics, University of Helsinki and Helsinki Institute of Physics, FIN-00014, Helsinki, Finland} 
\affiliation{University of Illinois, Urbana, Illinois 61801} 
\affiliation{The Johns Hopkins University, Baltimore, Maryland 21218} 
\affiliation{Institut f\"{u}r Experimentelle Kernphysik, Universit\"{a}t Karlsruhe, 76128 Karlsruhe, Germany} 
\affiliation{Center for High Energy Physics: Kyungpook National University, Daegu 702-701, Korea; Seoul National University, Seoul 151-742, Korea; Sungkyunkwan University, Suwon 440-746, Korea; Korea Institute of Science and Technology Information, Daejeon, 305-806, Korea; Chonnam National University, Gwangju, 500-757, Korea} 
\affiliation{Ernest Orlando Lawrence Berkeley National Laboratory, Berkeley, California 94720} 
\affiliation{University of Liverpool, Liverpool L69 7ZE, United Kingdom} 
\affiliation{University College London, London WC1E 6BT, United Kingdom} 
\affiliation{Centro de Investigaciones Energeticas Medioambientales y Tecnologicas, E-28040 Madrid, Spain} 
\affiliation{Massachusetts Institute of Technology, Cambridge, Massachusetts  02139} 
\affiliation{Institute of Particle Physics: McGill University, Montr\'{e}al, Qu\'{e}bec, Canada H3A~2T8; Simon Fraser University, Burnaby, British Columbia, Canada V5A~1S6; University of Toronto, Toronto, Ontario, Canada M5S~1A7; and TRIUMF, Vancouver, British Columbia, Canada V6T~2A3} 
\affiliation{University of Michigan, Ann Arbor, Michigan 48109} 
\affiliation{Michigan State University, East Lansing, Michigan  48824}
\affiliation{Institution for Theoretical and Experimental Physics, ITEP, Moscow 117259, Russia} 
\affiliation{University of New Mexico, Albuquerque, New Mexico 87131} 
\affiliation{Northwestern University, Evanston, Illinois  60208} 
\affiliation{The Ohio State University, Columbus, Ohio  43210} 
\affiliation{Okayama University, Okayama 700-8530, Japan} 
\affiliation{Osaka City University, Osaka 588, Japan} 
\affiliation{University of Oxford, Oxford OX1 3RH, United Kingdom} 
\affiliation{Istituto Nazionale di Fisica Nucleare, Sezione di Padova-Trento, $^z$University of Padova, I-35131 Padova, Italy} 
\affiliation{LPNHE, Universite Pierre et Marie Curie/IN2P3-CNRS, UMR7585, Paris, F-75252 France} 
\affiliation{University of Pennsylvania, Philadelphia, Pennsylvania 19104}
\affiliation{Istituto Nazionale di Fisica Nucleare Pisa, $^{aa}$University of Pisa, $^{bb}$University of Siena and $^{cc}$Scuola Normale Superiore, I-56127 Pisa, Italy} 
\affiliation{University of Pittsburgh, Pittsburgh, Pennsylvania 15260} 
\affiliation{Purdue University, West Lafayette, Indiana 47907} 
\affiliation{University of Rochester, Rochester, New York 14627} 
\affiliation{The Rockefeller University, New York, New York 10021} 
\affiliation{Istituto Nazionale di Fisica Nucleare, Sezione di Roma 1, $^{dd}$Sapienza Universit\`{a} di Roma, I-00185 Roma, Italy} 

\affiliation{Rutgers University, Piscataway, New Jersey 08855} 
\affiliation{Texas A\&M University, College Station, Texas 77843} 
\affiliation{Istituto Nazionale di Fisica Nucleare Trieste/Udine, I-34100 Trieste, $^{ee}$University of Trieste/Udine, I-33100 Udine, Italy} 
\affiliation{University of Tsukuba, Tsukuba, Ibaraki 305, Japan} 
\affiliation{Tufts University, Medford, Massachusetts 02155} 
\affiliation{Waseda University, Tokyo 169, Japan} 
\affiliation{Wayne State University, Detroit, Michigan  48201} 
\affiliation{University of Wisconsin, Madison, Wisconsin 53706} 
\affiliation{Yale University, New Haven, Connecticut 06520} 
\author{T.~Aaltonen}
\affiliation{Division of High Energy Physics, Department of Physics, University of Helsinki and Helsinki Institute of Physics, FIN-00014, Helsinki, Finland}
\author{J.~Adelman}
\affiliation{Enrico Fermi Institute, University of Chicago, Chicago, Illinois 60637}
\author{T.~Akimoto}
\affiliation{University of Tsukuba, Tsukuba, Ibaraki 305, Japan}
\author{B.~\'{A}lvarez~Gonz\'{a}lez$^t$}
\affiliation{Instituto de Fisica de Cantabria, CSIC-University of Cantabria, 39005 Santander, Spain}
\author{S.~Amerio$^z$}
\affiliation{Istituto Nazionale di Fisica Nucleare, Sezione di Padova-Trento, $^z$University of Padova, I-35131 Padova, Italy} 

\author{D.~Amidei}
\affiliation{University of Michigan, Ann Arbor, Michigan 48109}
\author{A.~Anastassov}
\affiliation{Northwestern University, Evanston, Illinois  60208}
\author{A.~Annovi}
\affiliation{Laboratori Nazionali di Frascati, Istituto Nazionale di Fisica Nucleare, I-00044 Frascati, Italy}
\author{J.~Antos}
\affiliation{Comenius University, 842 48 Bratislava, Slovakia; Institute of Experimental Physics, 040 01 Kosice, Slovakia}
\author{G.~Apollinari}
\affiliation{Fermi National Accelerator Laboratory, Batavia, Illinois 60510}
% ...

\collaboration{CDF Collaboration\footnote{With visitors from $^a$University of Massachusetts Amherst, Amherst, Massachusetts 01003,
$^b$Universiteit Antwerpen, B-2610 Antwerp, Belgium, 
$^c$University of Bristol, Bristol BS8 1TL, United Kingdom,
$^d$Chinese Academy of Sciences, Beijing 100864, China, 
$^e$Istituto Nazionale di Fisica Nucleare, Sezione di Cagliari, 09042 Monserrato (Cagliari), Italy,
$^f$University of California Irvine, Irvine, CA  92697, 
$^g$University of California Santa Cruz, Santa Cruz, CA  95064, 
$^h$Cornell University, Ithaca, NY  14853, 
$^i$University of Cyprus, Nicosia CY-1678, Cyprus, 
$^j$University College Dublin, Dublin 4, Ireland,
$^k$University of Edinburgh, Edinburgh EH9 3JZ, United Kingdom, 
$^l$University of Fukui, Fukui City, Fukui Prefecture, Japan 910-0017
$^m$Kinki University, Higashi-Osaka City, Japan 577-8502
$^n$Universidad Iberoamericana, Mexico D.F., Mexico,
$^o$University of Iowa, Iowa City, IA  52242,
$^p$Queen Mary, University of London, London, E1 4NS, England,
$^q$University of Manchester, Manchester M13 9PL, England, 
$^r$Nagasaki Institute of Applied Science, Nagasaki, Japan, 
$^s$University of Notre Dame, Notre Dame, IN 46556,
$^t$University de Oviedo, E-33007 Oviedo, Spain, 
$^u$Texas Tech University, Lubbock, TX  79609, 
$^v$IFIC(CSIC-Universitat de Valencia), 46071 Valencia, Spain,
$^w$University of Virginia, Charlottesville, VA  22904,
$^x$Bergische Universit\"at Wuppertal, 42097 Wuppertal, Germany,
$^{ff}$On leave from J.~Stefan Institute, Ljubljana, Slovenia, 
}}
\noaffiliation

\date{\today}

\begin{abstract}
We present three measurements of the top-quark mass in the lepton plus jets channel with approximately \lumi\ of integrated luminosity collected with the CDF II detector
using quantities with minimal dependence on the jet energy scale. One measurement
exploits the transverse decay length of \Bot-tagged jets to determine a top-quark mass of \lxyRes, and another the transverse momentum of electrons and muons from \W-boson decays to determine a top-quark mass of \lepPtRes. These quantities are combined in a third, simultaneous mass measurement to determine a top-quark mass of \combRes. 
\end{abstract}


\maketitle

\section{Introduction \label{sec:intro}}
 
An accurate knowledge of the top-quark mass is important. Combined with other standard model parameters, the top-quark mass can be used to constrain the expected standard model Higgs mass. The most precise constraints place an upper bound on the Higgs mass of $154\ \textnormal{GeV}/c^2$ \cite{bib:lep_EW_higgs} at the 95\% confidence level. Further, since the top quark will be produced in copious quantities at the LHC, its mass can serve as a benchmark for \Bot-jet energy calibrations, which are otherwise difficult to study in data. 

Historically, top-quark mass measurements have been limited by uncertainties in the simulation of jet energy measurements. For example, in the CDF Run I measurement \cite{bib:runI_top_mass} the top-quark mass systematic uncertainty was $4.9\ \textnormal{GeV}/c^2$, of which the contribution from the uncertainty on jet energy measurements was $4.4\ \textnormal{GeV}/c^2$. In comparison, for a modern CDF Run II measurement \cite{bib:runII_top_mass}, the systematic uncertainty is $1.2\ \textnormal{GeV}/c^2$, of which the contribution from jet energy uncertainties is $0.7\ \textnormal{GeV}/c^2$. This dramatic improvement in the jet energy uncertainty is made possible by exploiting the hadronic \W-boson decay. By constraining the reconstructed \W-boson mass to agree between data and simulation, the average simulated jet energy bias is determined and applied to all jets in the event. In this technique, widely used in Run II top-quark mass measurements, most of the jet energy uncertainty becomes a statistical uncertainty rather than a systematic uncertainty. The remaining systematic uncertainty results from the assumption that simulation is biased by the same amount, on average, for the \Bot-jets as it is for the \W-boson jets that are used for the calibration.

While jet energy uncertainties are no longer as large as they once were, they still represent a significant fraction of the total uncertainty on the top-quark mass. Further, this dramatic improvement in the world average mass resolution depends upon the reliability of a single technique. Thus, it is desirable to develop independent methods to measure the top-quark mass as a cross-check. A novel technique has been proposed for measuring the top-quark mass in a manner that is almost completely independent of the calorimeter using the measured transverse distance \cite{bib:CDF_coords} that  \Bot-hadrons from the top quarks travel before decaying (Lxy) \cite{bib:PRD_lxy_proposal}. The transverse momenta of the top-quark decay products depend approximately linearly on the top-quark mass. In turn, this means that the lifetime and decay length of the \Bot-hadrons depend approximately linearly on the top-quark mass. When this measurement was performed using 695 \ipb\ of integrated luminosity \cite{bib:lxy_first_measurement}, it became the first top-quark mass measurement to be mostly independent of calorimeter-based uncertainties, however it was also very statistically limited. Besides tripling the integrated luminosity with respect to the previous measurement, we also improve the statistical resolution by incorporating more information about the event. Similarly to \Bot-hadrons, the transverse momenta ($p_T$) of leptons from \W-boson decays also depend linearly on the top-quark mass, and are mostly independent of the calorimeter jet energy scale. Since the momentum of the leptons is mostly uncorrelated to that of the \Bot-quarks, this is complementary information, and it is an ideal variable to add to the measurement, as has been previously proposed  \cite{bib:PRD_lxy_proposal} \cite{bib:Nikos_lepPt}.     
    
\begin{figure*}
  \centering
  \subfigure[] { 
    \includegraphics[height=2.1in]{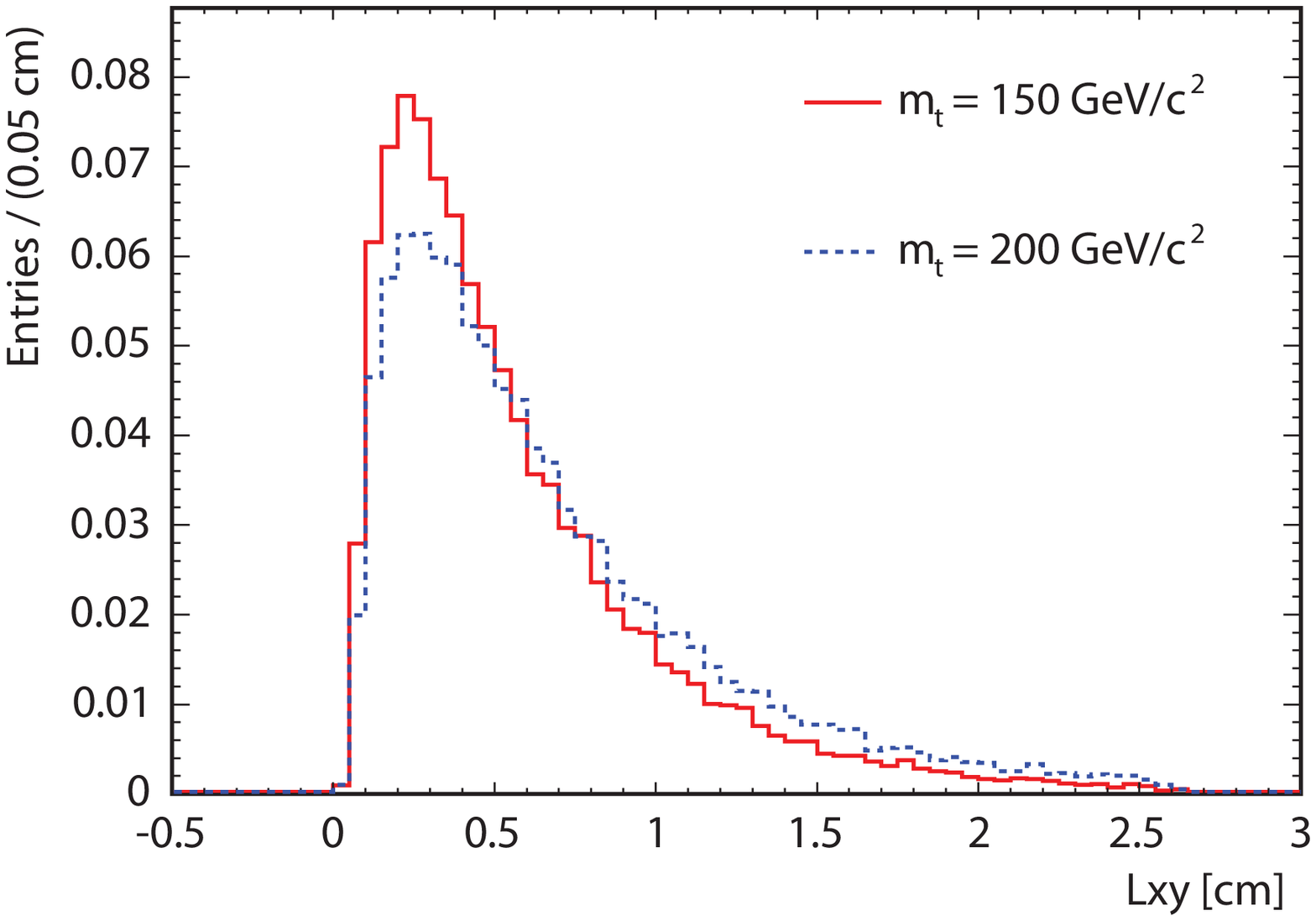}
  }
  \subfigure[]  {
     \includegraphics[height=2.1in]{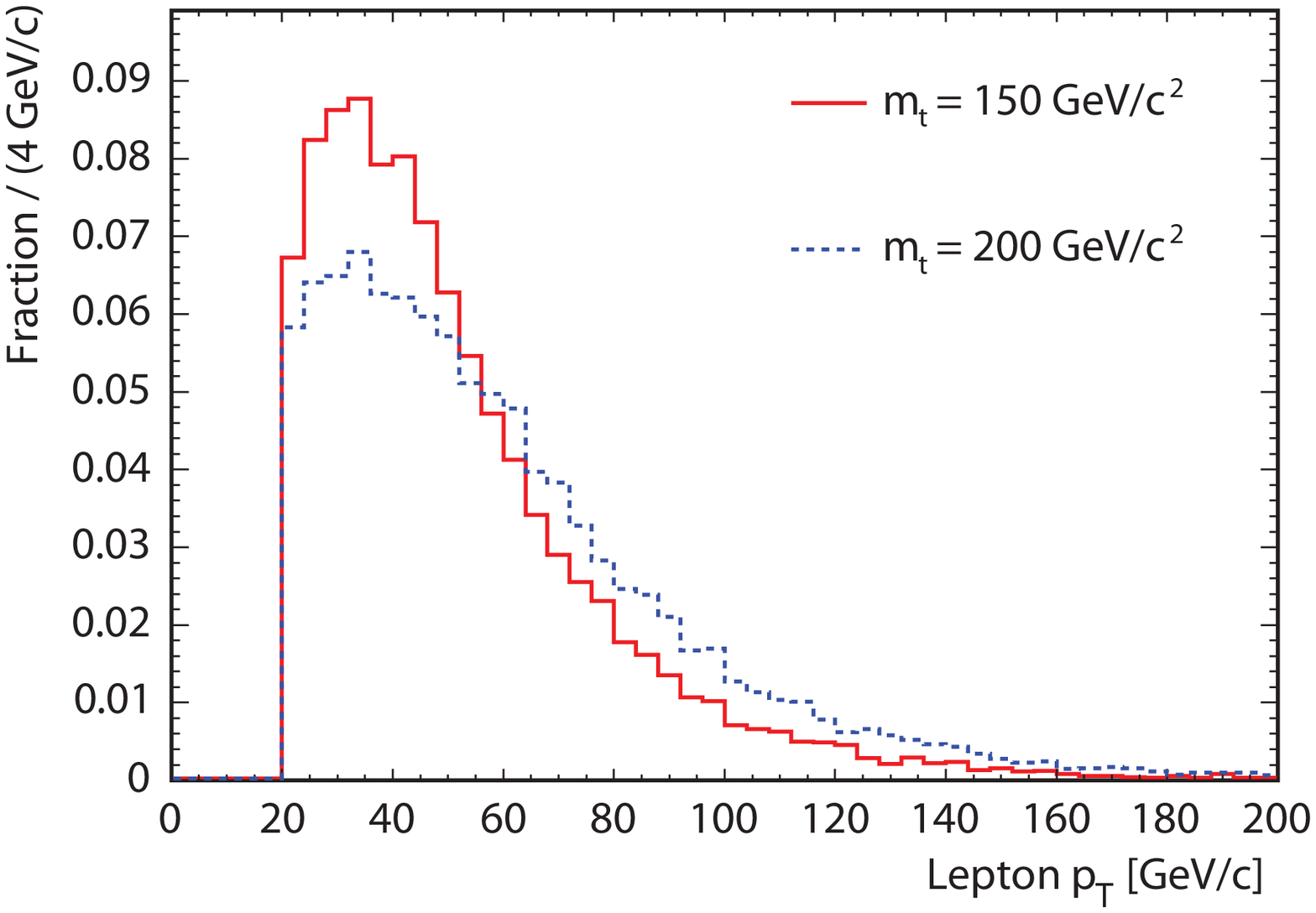}
   }
  \caption{Distributions of our measurement variables for simulated \ttbar\ events after passing our event selection for top-quark masses separated by $50\ \textnormal{GeV}/c^2$. These distributions are normalized to unit area.}\label{fig:dists}
\end{figure*}

                                                   
In this paper we present measurements of the top-quark mass using both the lepton transverse momentum and the mean decay length variables. Both measurements are performed upon the same events which pass an event selection that is designed to isolate \ttbar\ events where the \W-boson from one top-quark decays to two quark jets, and the other decays leptonically to an electron or a muon plus a neutrino (the lepton plus jets channel). This decay channel was chosen because it occurs more often than any other final state that can be isolated with a high signal purity. We compute the mean Lxy of the leading two \Bot-tagged jets and the mean transverse momentum of the leptons identified in the events. We measure the top-quark mass through comparisons with the mean Lxy and mean lepton \pT\ from analysis simulations (``pseudoexperiments") performed for a variety of top-quark mass hypotheses. To illustrate the dependence of these variables on the top-quark mass, the expected distributions for \ttbar\ signal events passing our event selection are shown in Figure~\ref{fig:dists}. It should be noted that our measurements are sensitive to very different event characteristics than typical mass measurements, and thus require unique treatments. In particular, it is more important for us to correctly model the boost of the top quarks than for other measurements. Thus, we reweight our simulated signal events to a more accurate parton distribution function. Further, the Lxy values that are measured in our simulation are sensitive to a variety of possible modeling inaccuracies. We determine correction factors to be applied to the simulated Lxy values by comparing the Lxy measurements between data and simulation for \bbbar\ events, parameterized as a function of the jet energies. 

Our measurements will be mostly independent of calorimeter-based jet energy uncertainties. However, some small dependence will remain because a loose jet energy cut is applied when selecting events, as will be explained below. Further, this parameterization of the Lxy correction as a function of jet energy introduces a jet energy uncertainty to the measurement. In order to minimize this latter effect, we develop and apply an algorithm to measure jet energies for our Lxy correction parameterization that is based upon charged particle tracking.


This paper is organized as follows. We first present the relevant parts of the CDF detector in Section \ref{sec:detector}, and the event selection used in this analysis in Section \ref{sec:evt_selection}. We then explain the procedures for determining the normalizations as well as the shapes of the Lxy and lepton $p_T$\ distributions for our backgrounds. In Section \ref{sec:corrections} we explain how we calibrate our signal distributions. In Section~\ref{sec:method} we present the procedures we apply to extract the top-quark mass from our final Lxy and lepton momentum distributions and present our results. Finally, in Section~\ref{sec:systs} we explain in detail how our systematic uncertainties are estimated. We conclude with some projections of the future potential for this type of measurement. 

\section{The CDF II Detector \label{sec:detector}}
We describe the most important parts of the CDF detector for this analysis. A
more complete description can be found elsewhere \cite{bib:CDF_detector}. The CDF detector consists of cylindrically symmetric layers of hardware, each designed to perform specific functions. The CDF tracking system consists of a silicon tracker, inside of a
wire drift chamber, immersed in a 1.4 T magnetic field produced by a superconducting solenoid. Calorimeters encircle the tracking system and consist of alternating layers of scintillator and absorber. The calorimeter system is split into an inner calorimeter that is designed to absorb and measure the energies of photons and electrons, and an outer calorimeter that is designed to absorb and measure the energies of hadrons. 

The wire drift chamber consists of 96 wire planes grouped into eight superlayers covering the radial region between 0.40 and 1.37 m from the beam axis, and less than 1.0 in pseudorapidity. The superlayers alternate between axial and $\pm 2^\circ$\ stereo angle to provide both axial and longitudinal hit information. The inner tracker consists of eight layers of double
sided silicon, covering the radial range between 1.35 and 29 cm, and less than 2.0 in
pseudorapidity. Again, to provide longitudinal hit information, some layers contain double-sided silicon that combines axially oriented strips with strips oriented either at $90^\circ$\ or $1.2^\circ$\ relative to the axial direction. The silicon tracker is vital for
vertexing, providing a two-dimensional impact parameter resolution of about 40 $\mu$m
for isolated tracks with high transverse momentum, including uncertainties on both the track trajectory and the location of the collision vertex. The
wire tracker is vital for the lepton measurements in this analysis, providing a
transverse momentum resolution of about 5\% for 50 GeV/$c$\ muons.  Leptons and \Bot-jets
are reliably identified out to a pseudorapidity of about 1.0, after which the efficiency drops rapidly due to tracks falling outside of the range of full tracker coverage. Muon candidates are identified by matching tracks in the inner tracking detector with track stubs in the outer muon drift chambers.
Electron candidates are identified by matching tracks in the inner tracking detector with showers in the electromagnetic calorimeter. An energy resolution of about 3\% is achieved for electrons with 50 GeV of transverse energy. 

The Tevatron proton-antiproton bunches cross one another at a rate of 2.5 MHz. During each crossing one or more collisions are likely to occur. In order to reduce this flow of information to a manageable level and to select collision events of interest to particular analyses, CDF employs a three level triggering system to sequentially reject events that are less relevant. The final event stream is read out at a rate of roughly 100 Hz. Details of the triggering system are provided in \cite{bib:CDF_detector}.

\section{Event Selection \label{sec:evt_selection}} 

In comparison to the previous publication using only Lxy \cite{bib:lxy_first_measurement}, we tighten the event selection in the analyses presented here to reduce the systematic uncertainties. The statistical sensitivities of both the lepton momentum and the decay length measurements depend linearly on the fraction of events in which top quarks are produced\cite{bib:mean_exp}, and scale as the square root of the number of events selected. This tightened selection improves the former and worsens the latter, and has little impact on the final statistical sensitivity.

The data used in this analysis were collected between March 2002 and May 2007,
and correspond to an integrated luminosity of \lumi. Events passing the full
trigger and event selections described below are studied to determine the expected event counts
and uncertainties for each signal and background type. Top quarks almost always decay to \W\Bot. Our selection criteria are designed to accept events where one \W-boson decays to an electron or muon plus neutrino and the other decays to two jets. We start from a triggering stream that requires one electron (muon) to have transverse energy (momentum) greater than 18 GeV (GeV/$c$); these requirements will later be tightened slightly. Once events are accepted by the trigger, they are saved, reconstructed, and studied in greater detail. Calorimeter towers are clustered together within a cone of radius $R = \sqrt{(\phi_{tow}-\phi_{jet})^2+(\eta_{tow}-\eta_{jet})^2} = 0.4$ \cite{bib:newer_CDF_JES_note} to form jets. At least three jets must be found with $|\eta| < 2.0$, and transverse energy greater than 20 GeV after correcting for multiple interactions, calorimeter response, noise, and other non-uniformities. No attempt is made to correct for interactions from spectator partons (``underlying event") or out-of-cone effects  \cite{bib:newer_CDF_JES_note}, but systematic uncertainties are assigned for these effects. To account for the neutrino and suppress the QCD background, a quantity called the missing transverse energy is used, which is defined as the transverse component of the four-momentum vector that is needed to  conserve momentum in the event \cite{bib:MET_explic}. The missing transverse energy in the event must be greater than 20 GeV. Additionally, an electron (muon) must be identified with transverse energy (momentum) greater than 20 $\textnormal{GeV}$ ($\textnormal{GeV}/c$). Electrons are identified from a track pointing at a cluster in the calorimeter which matches the expected shape profile. Most of the energy of this cluster is required to be confined in the electromagnetic calorimeter, the track momentum is required to agree with the measured calorimeter energy to within a factor of two, and if the track is consistent with the electron having originated from a photon conversion, the electron candidate is vetoed. Muons are formed from tracks in the muon chambers which are matched to tracks found in the inner tracking system. The calorimeter energy deposits along the muon trajectory must be consistent with that of a minimum ionizing particle. Calorimeter isolation is based upon the fraction of the lepton's energy ($f_{iso}$) in a cone of radius R=0.4 centered on the lepton,  excluding the energy in the calorimeters from the lepton itself. Both electrons and muons must satisfy $f_{iso}~<~0.1$. This cut eliminates most jets that are mistakenly identified as leptons (``fake leptons") as well as real leptons which result from \Bot-hadron decays. Further, at least one collision vertex must be reconstructed from tracks in the event, and the track of the lepton must pass within 3 cm along the beam axis of the highest momentum vertex to minimize contamination from multiple interactions and tracking errors. One electron or muon is required to pass these cuts, but to suppress events from the dilepton channel the event is vetoed if any other leptons are found passing a much looser set of cuts which also allow non-isolated and forward leptons. Finally, one or more jets must be identified (tagged) as originating from a \Bot-quark, as explained below. In the case where only three jets pass our selection, two of them must be tagged in order to reduce the large \W+jets and non-\W\ QCD backgrounds in this kinematic region. 


Jets containing \Bot-quarks are identified using a tagging algorithm known as SecVtx \cite{bib:SecVtx_ref}, which relies upon the long lifetime of hadrons originating from \Bot-quarks. This algorithm attempts to construct a secondary vertex using tracks that are likely to have originated from a \Bot-hadron decay. If a vertex can be found which is significantly displaced from the primary vertex, then the jet is tagged. Specifically, the SecVtx algorithm selects tracks associated with the jet that are displaced from the fitted primary collision vertex, and that are well resolved in both the silicon and the outer tracking chamber. As a first pass the algorithm uses relatively less stringent cuts on track selection, and attempts to fit at least three tracks into a displaced vertex. If no vertex is found the tracking requirements are tightened significantly and the algorithm searches for a two-track vertex. Further cuts are applied to eliminate tracks that are likely to originate from material interactions or long lived strange particles. If a secondary vertex is found the jet is tagged as a \Bot-jet if the transverse distance from the primary vertex projected onto the jet direction (Lxy) and its uncertainty ($\sigma$) satisfy $\textnormal{Lxy} / \sigma~>~ 7.5$.  It should be noted that charmed daughters of the \Bot-hadron are also likely to travel a significant distance before decaying. The SecVtx algorithm is deliberately designed to be loose enough to attach some tracks from these tertiary decays into one ``pseudovertex" at a position that is averaged between two real vertices. Since the boosts of the charm hadrons depend on the boost of the \Bot-hadron, this extra information does not dilute our mass resolution.

\section{Sample composition \label{sec:sample_comp}}

For this analysis we will need to know the Lxy and lepton $p_T$\ distributions for each of our signal and background samples, as well as their relative normalizations after full event selection is enforced. Specifically, we normalize our signal and background distributions using the results of a cross section measurement that was performed on the same data using identical event selection. The procedures for this cross section measurement are briefly described in subsection~\ref{sec:cross_section}. The procedures that are used to determine our Lxy and lepton \pT\ shapes for the backgrounds are deferred to Section~\ref{sec:bkg_shapes}.

\subsection{Sample normalization \label{sec:cross_section}}

In this section we give a brief outline of how the cross section measurement is performed to determine our signal and background normalizations along with the associated uncertainties. Further information can be found in the publications of similar cross section measurements \cite{bib:lpj_old_cross_section} \cite{bib:lpj_old_cross_section_detailed}, or in the Ph.D. thesis \cite{bib:Sherman_thesis} which explains the technique we apply in detail. For this measurement technique we begin by determining the expected backgrounds assuming the expected standard model \ttbar\ cross section. The \ttbar\ cross section will then be revised and the backgrounds redetermined in an iterative manner until the number of expected events matches the observed event counts both before and after \Bot-tagging. 

First we determine the numbers of
events for some of the rarest processes from simulation. Backgrounds modeled from simulation 
include single top production as well as the electroweak diboson (\W\W, \W\Z, \Z\Z) and \Z\ plus jets final states. The diboson events are simulated using
\textsc{pythia} version 6.216 \cite{bib:Pythia_ref}. The single
top and \Z\ plus jets samples are simulated using other programs (MadEvent \cite{bib:Madevent_ref} for single top, and \textsc{alpgen} version 2.10 prime \cite{bib:Alpgen_ref} for \Z\ plus jets). Hadronic showers are simulated using \textsc{pythia}. For each sample, the event count is determined and scaled according to the theoretical cross section, branching ratio, and detector and trigger acceptances. The \ttbar\ signal is also simulated using \textsc{pythia}. Its cross section is initially set equal to its standard model expectation. For all of these samples, EvtGen \cite{bib:EvtGen_ref} is used to determine the lifetimes and masses of the various \Bot-hadron species within jets. A further scaling is applied to correct the \Bot-tagging efficiency modeling based upon data-driven studies~\cite{bib:lpj_old_cross_section_detailed}.

Since fake leptons are difficult to simulate accurately, the non-\W\ QCD contribution is determined from data. Missing transverse energy templates are made for \ttbar, \W\ plus jet, and QCD distributions, and used in a fit to determine the QCD normalization.  Templates for the fake electrons are filled from events where isolated electron candidates are selected and required to have a calorimeter shower profile that is consistent with QCD fakes rather than true electrons. For muons the standard cuts are kept except that the isolation cut is inverted to require greater than 20\% isolation instead of less than 10\%. Different binnings, fit ranges, and cuts are applied and the differences in the results are taken as a systematic uncertainty. This fit is done separately in the tagged and pretagged cases.

The remainder of the observed pretagged events are all taken to originate from \W\ plus jets. The \W\ plus jets sample is simulated using \textsc{alpgen} version 2.10 prime \cite{bib:Alpgen_ref}, and the resulting partons are showered using \textsc{pythia}. This sample is simulated in various bins of jet multiplicity for the \W\bbbar, \W\ccbar, \W$c$, and \W\ plus light flavor final states. These samples are then weighted by their theoretical cross sections and combined. It is important to make sure that the proper number of heavy flavor jets are simulated before \Bot-tagging is applied. Pythia showering will double count heavy flavor production from the \textsc{alpgen} simulation, so this overlap must be removed. Further, since the samples are only generated at leading order, a correction is needed to reweight the heavy flavor fraction. This correction is taken from comparisons between data and simulation in related samples with higher statistics \cite{bib:lpj_old_cross_section_detailed}.

By construction, this procedure yields pretagged background and signal normalizations that exactly match the observed number of data events. After \Bot-tagging, however, fewer events were predicted than were observed, and this discrepancy is attributed to \ttbar\ events. Thus, the signal cross section was increased and the analysis was repeated in an iterative fashion. The \ttbar\ cross section at which we found agreement was $8.2 \pm 0.7\ pb$ (ignoring luminosity uncertainties). The resulting event counts passing full event selection and tagging requirements are shown in Table~\ref{MII_res_table}, along with the number of jets in these events which were \Bot-tagged and included in our Lxy analysis.

\begin{table}[th]
\begin{center}
\caption{Estimated signal and background contributions after full event selection.}\label{MII_res_table}
\begin{ruledtabular}
\begin{tabular}{ccc}
Source & Events & Recorded \Bot-Tags  \\
\hline
\W\bbbar\ &  $25.4 \pm 7.0$ &  $37.5 \pm 9.7$\\
\W\ccbar\ or \W$c$ &  $13.9 \pm 4.6$ &  $15.9 \pm 4.7$\\
\W plus light flavor &  $16.9 \pm 3.7$ &  $17.9 \pm 3.7$\\
non-\W QCD &  $18.8 \pm 12.7$ &  $20.2 \pm 13.2$\\
Electroweak &  $9.0 \pm 0.4$ &  $11.3 \pm 0.5$\\
Single Top &  $8.4 \pm 0.4$ &  $13.4 \pm 0.7$\\
\ttbar\ &  $478.3 \pm 40.3$ &  $659.3 \pm 45.5$\\
\hline
Total &  $570.8 \pm 44.3$ &  $775.5 \pm 50.2$\\
\end{tabular}
\end{ruledtabular}
\end{center}
\end{table}

\subsection{Background Lxy and lepton $p_T$\ shapes \label{sec:bkg_shapes}}

The dominant backgrounds for our analysis are \W\ plus heavy flavor (\Bot\ and \Charm\ jets), \W\ plus light flavor mistags, and
non-\W\ QCD events. Along with single top and diboson events, these distributions and the signal account for about 99\% of
events passing selection. The remaining events come from the \Z\ plus jets background, for which the related \W\ plus jets background Lxy and lepton \pT\ distributions were used.  

Single top samples (with masses $m_t = 165,\ 170,\ 175,$\ and\ $180\ \textnormal{GeV}/c^2$) were generated in MadEvent and decayed in \textsc{pythia}, in both the s- and t- channels. Results for the s- and t- channels were combined, weighted according to their expected theoretical cross sections, and the final decay length and lepton momentum distributions were fit to Gaussian plus exponential distributions. The trends in the fit parameters were extrapolated to other mass points, and were used to generate new Lxy and lepton $p_T$ distributions for each mass hypothesis. The \W\ plus jets background is selected in exactly the same way as for the cross
section measurement, as is the background for non-\W\ QCD electrons. For non-\W\ QCD background in the muon channel, instead of using non-isolated muons, we selected muon candidates with high energy deposition in the calorimeter, or that were associated with tracks that were highly displaced from the collision vertex. These cuts were chosen based on studies suggesting that they produced a
minimum of bias in the lepton $p_T$ distribution. 

\begin{figure*}
\centerline{
  \subfigure[] {\includegraphics[height=2.5in]{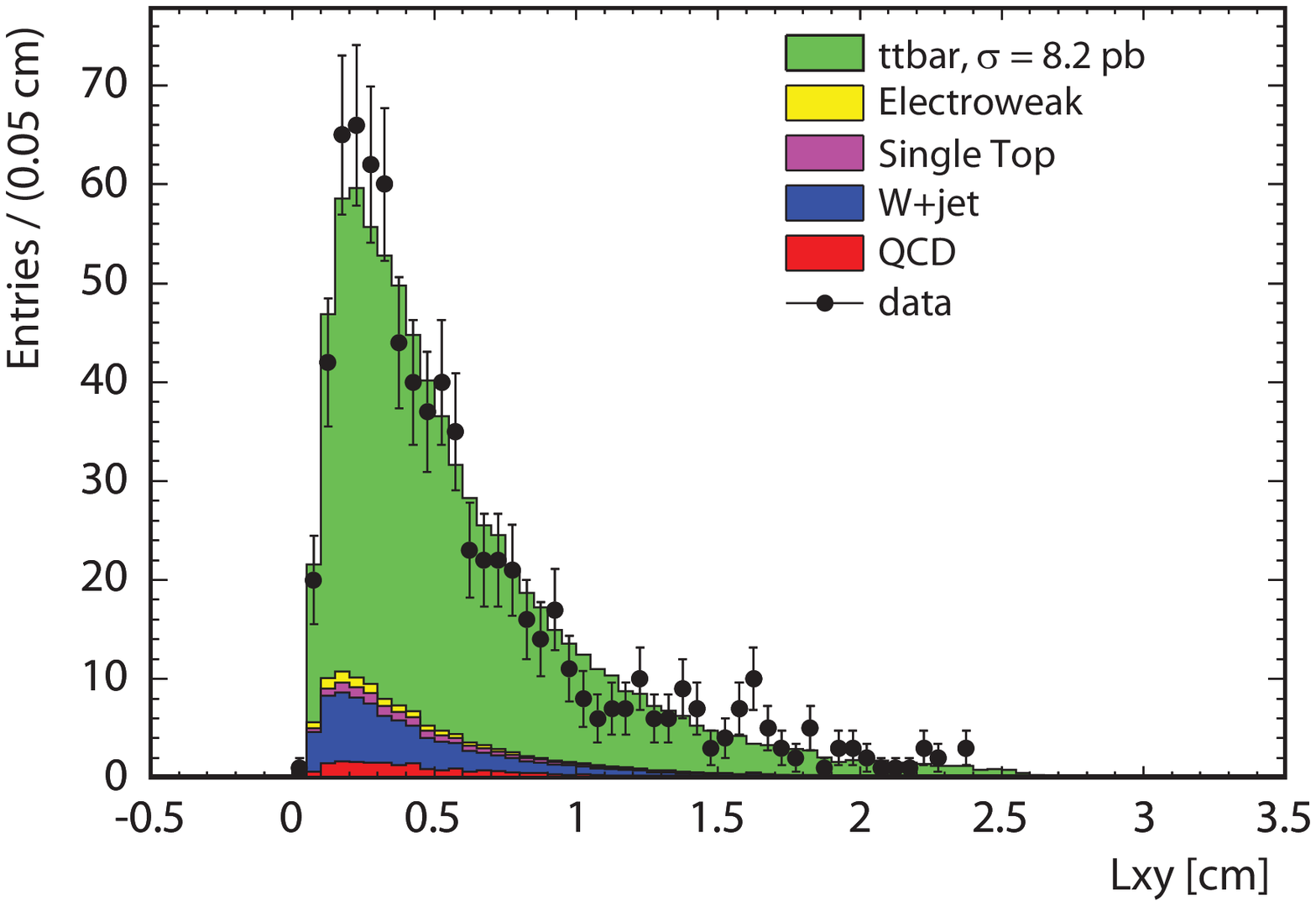}}
  \subfigure[] {\includegraphics[height=2.5in]{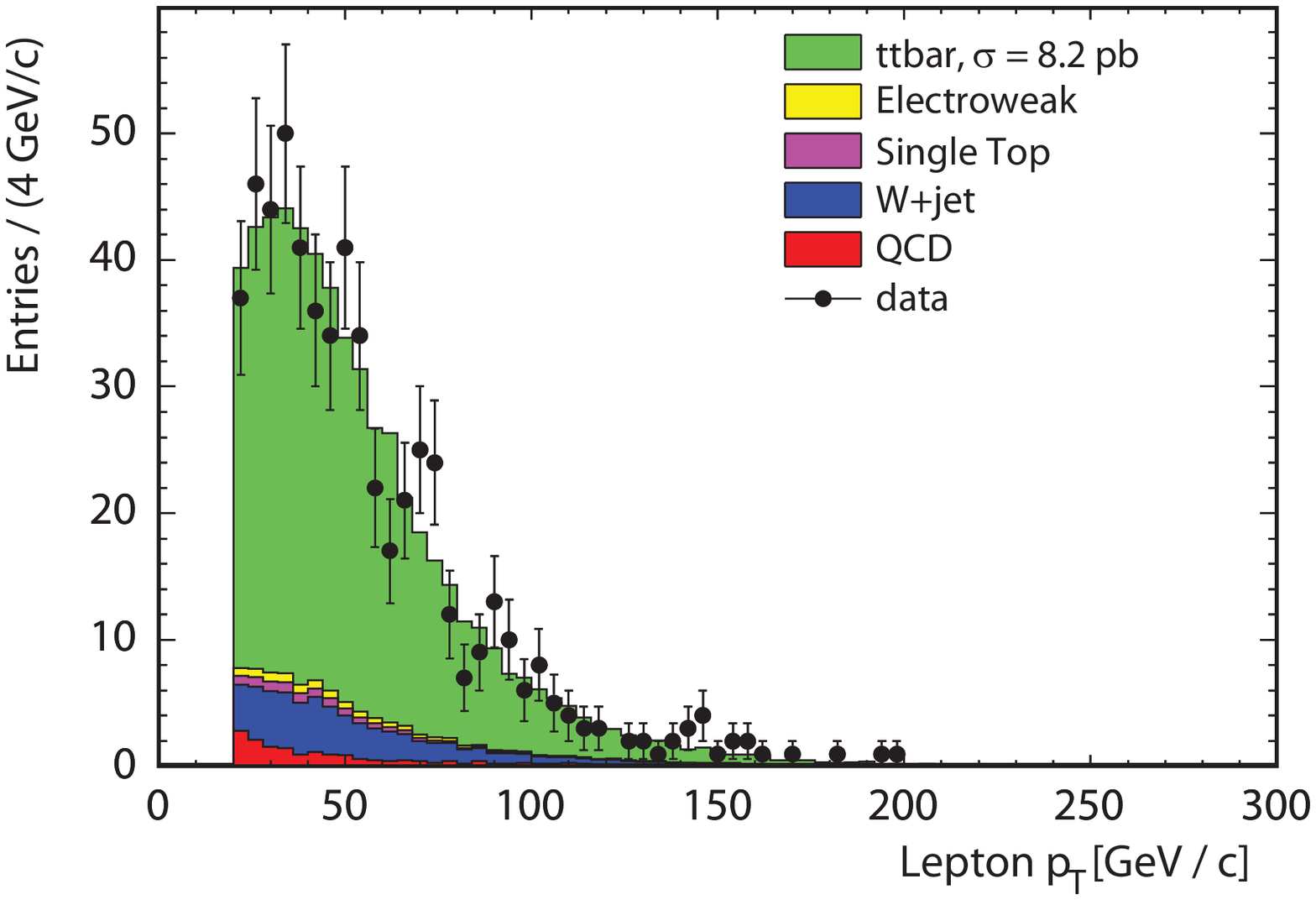}}
}
\caption{Signal, background, and data for the Lxy and lepton $p_T$\ distributions passing full event selection under hypothesized top-quark masses that are close to the measured results. The left plot is for the Lxy measurement, using top-quark mass $168\ \textnormal{GeV}/c^2$, and the right plot is for the lepton $p_T$  measurement, using top-quark mass $173\ \textnormal{GeV}/c^2$.\label{fig:meas_stacks}}
\end{figure*}

\begin{figure*}
\centerline{
  \subfigure[] {\includegraphics[height=2.5in]{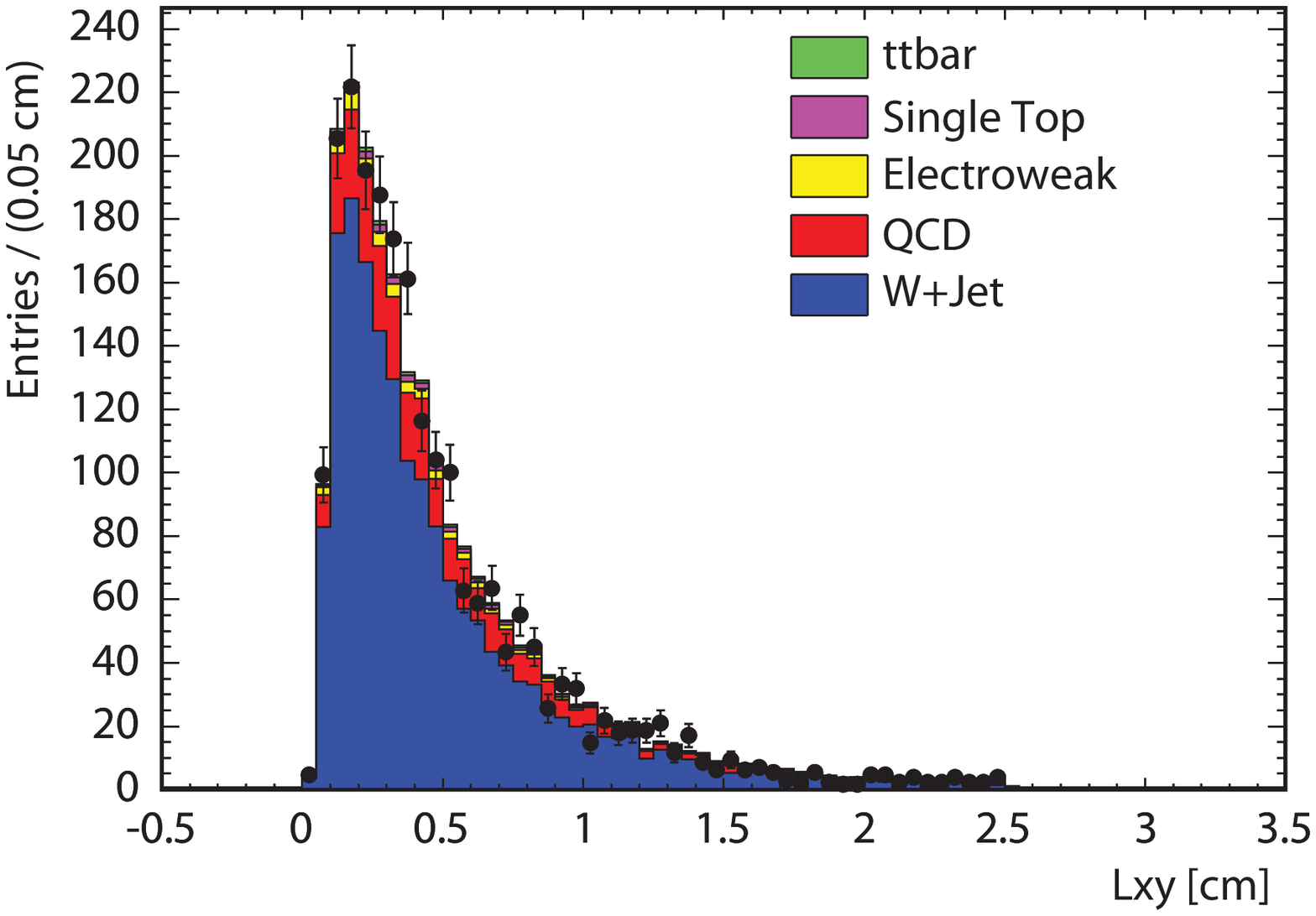}}
  \subfigure[] {\includegraphics[height=2.5in]{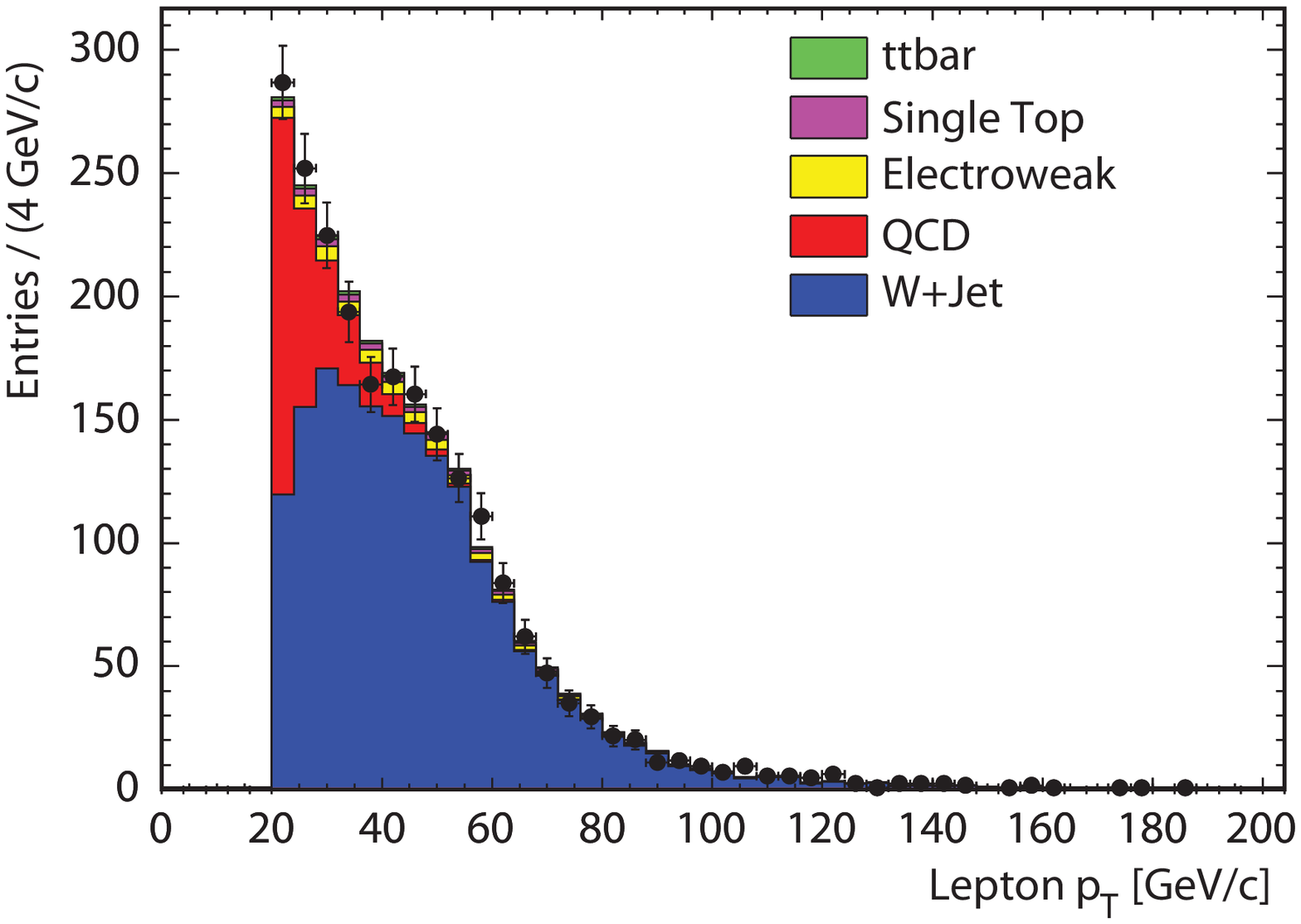}}
}
\caption{Background prediction compared with data (black points) in the one-jet control region for Lxy (left) and lepton $p_T$ (right).\label{fig:onejet_cross_check}}
\end{figure*}

\begin{figure*}
\centerline{
  \subfigure[] {\includegraphics[height=2.5in]{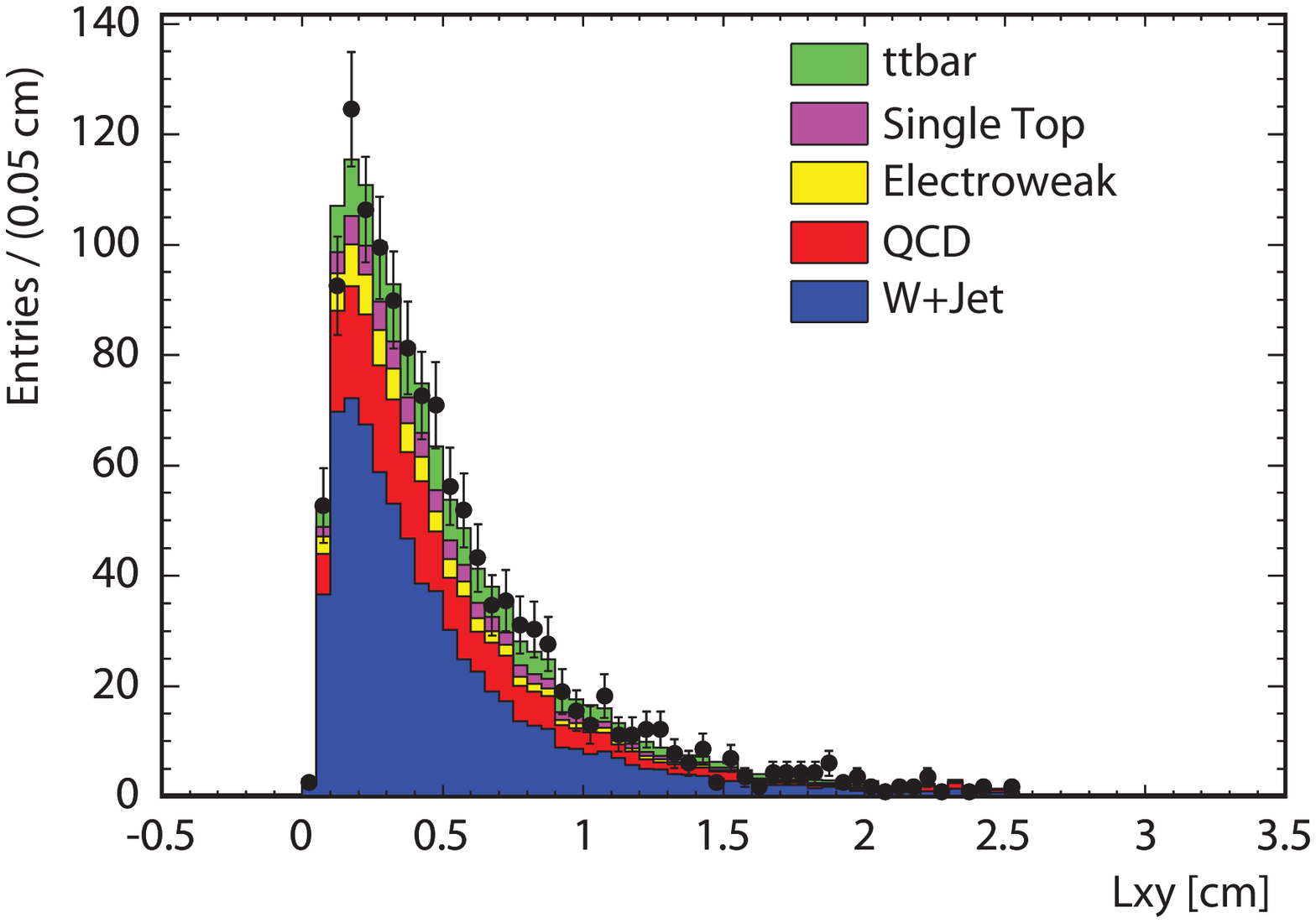}}
  \subfigure[] {\includegraphics[height=2.5in]{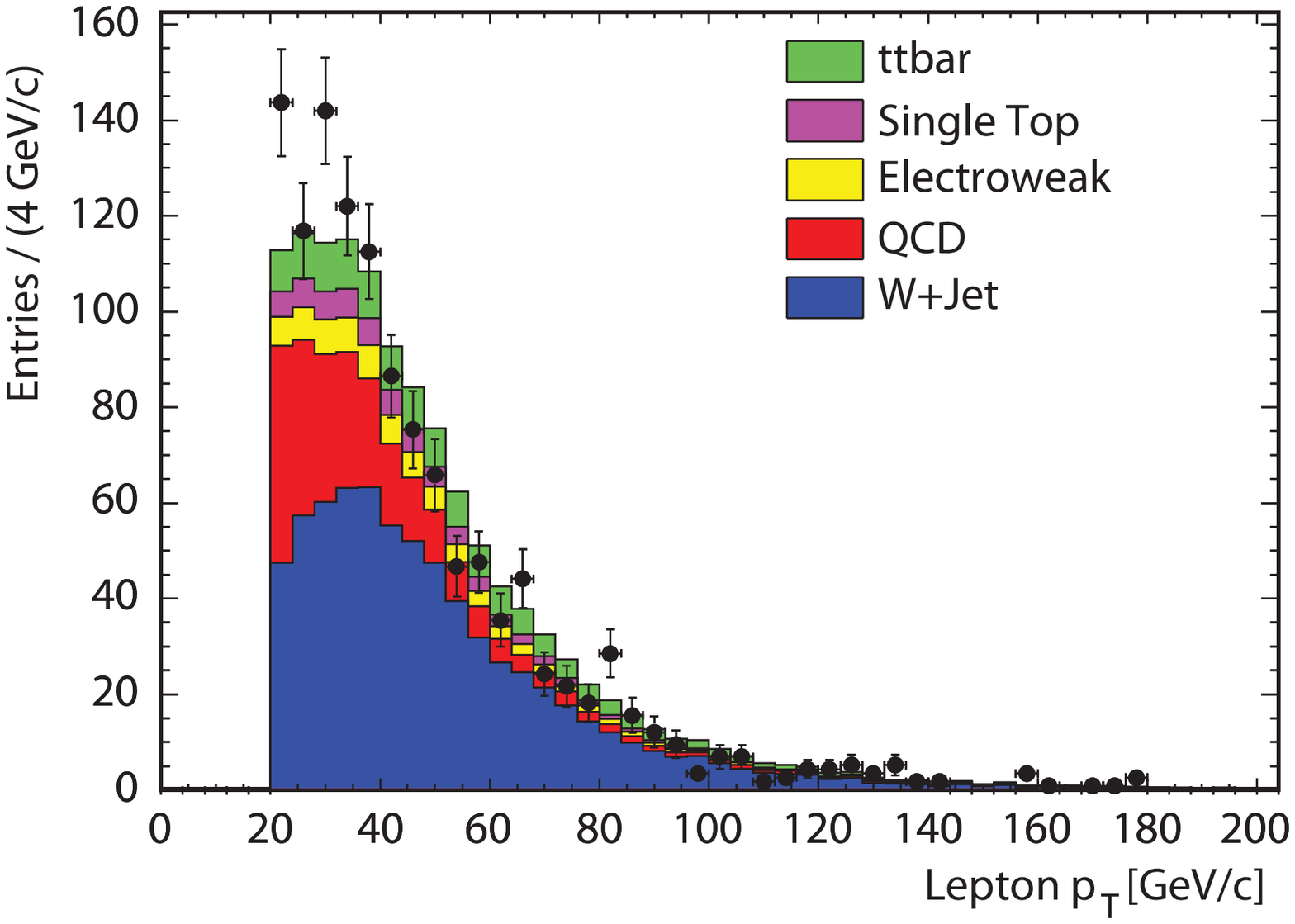}}
}
\caption{Background prediction compared with data (black points) in the two-jet control region for Lxy (left) and lepton $p_T$ (right).\label
{fig:twojet_cross_check}}
\end{figure*}

Our final Lxy and lepton $p_T$\ distributions are shown in Figure
~\ref{fig:meas_stacks} for events passing full selection. As cross-checks the same distributions in the background dominated one and two jet bins are shown in Figures ~\ref{fig:onejet_cross_check} and ~\ref{fig:twojet_cross_check}. These cross-check samples are used in evaluating the background based systematic uncertainty as described in Section~\ref{sec:systs}.



\section{Corrections to the signal simulation\label{sec:corrections}}

Our \ttbar\ events are simulated in \textsc{pythia}\ for various hypothetical top-quark mass values. A number of corrections are applied to the simulated results as discussed below. In Section~\ref{sec:PDFCors} we will discuss corrections that are needed to account for simulated parton distribution function inaccuracies. It is also necessary to correct the simulated Lxy measurements to match observations from data. An overview for this procedure is given in Section~\ref{sec:lxy_weightings}. We will parameterize the Lxy correction as a function of the jet transverse energy as measured using tracking. The procedure for this will be discussed in Section~\ref{sec:track_en}. We will summarize the complete decay length calibration procedure in Section~\ref{sec:final_lxy_calib}.

\subsection{Parton distribution functions \label{sec:PDFCors}}

For this analysis it is vital to have accurate models of the lepton and jet boosts. Since these quantities depend on the energy of the colliding partons, an accurate modeling of parton distribution functions (PDFs) for particles within the proton is very important. The \ttbar\ samples were generated using the leading order CTEQ5L parton distribution function~\cite{bib:CTEQ5L_ref}.  One drawback of this PDF is that it underestimates the rate of \ttbar\ production by gluon fusion (5\% in simulation observed versus about 15\% expected \cite{bib:TopReview}). Since gluon fusion produces events with slightly smaller boosts on average, this will bias the analysis. In addition, since the CTEQ5L PDF overestimates the fraction of the proton momentum carried by colliding quarks, the \ttbar\ events produced within the quark-antiquark annihilation channel have artificially high boosts. To compensate for this effect, different parton distribution functions were studied. The Les Houches Accord PDF Interface \cite{bib:LHAPDF_ref} was used to determine the probabilities for each parton involved in the collision to have the generated momentum for a given PDF. One can then reweight the simulated events to construct distributions appropriate to a different PDF as explained in \cite{bib:PDFReweightRef}. We use this prescription to reweight all of our \ttbar\ events to the next-to-leading order CTEQ6M~\cite{bib:CTEQ6M_ref} parton distribution function for gluon fusion and quark annihilation events separately. A further weighting is then applied to gluon fusion events to scale the gluon fusion fraction to the value expected for the sample's top-quark mass (20\% for $m_t = 150\ \textnormal{GeV}/c^2$, 10\% for $m_t = 200\ \textnormal{GeV}/c^2$). These combined reweightings result in new distributions to be used in our mass measurement and lead to a $1.7\ \textnormal{GeV}/c^2$ shift in the top-quark mass for the lepton $p_T$ analysis and a $0.9\ \textnormal{GeV}/c^2$ shift for the Lxy analysis, relative to the values obtained using CTEQ5L. A similar prescription is used to reweight to other PDFs and gluon fractions to evaluate systematic uncertainties, as will be explained in Section~\ref{sec:systs}.

\subsection{Lxy calibration strategy\label{sec:lxy_weightings}}

A number of effects may bias the decay length measurement in simulation. Inaccuracies in the EvtGen \cite{bib:EvtGen_ref} values for the hadron lifetime or in the simulated production fractions would have a direct impact. Similarly, inaccuracies in the \textsc{pythia} fragmentation model would lead to the wrong boost, and thus the wrong average decay length, of the \Bot-hadrons. Entirely different problems may arise from any inaccuracies in the modeling of the tracking system, which could lead to biases in the vertexing results. Our approach is to calibrate the simulation directly to the data to compensate for all of these biases simultaneously. Systematic uncertainties on this calibration will be discussed in Section~\ref{sec:Lxy_uncertainties}. We select our calibration sample so that the tagged jets will be almost exclusively \Bot-jets. We select dijet events where the jets are required to open at a wide angle with $\Delta\phi > 2.0$. Both jets must be \Bot-tagged, and one of the jets must contain a well resolved muon with at least $9.5\ \textnormal{GeV}/c$ transverse momentum. We also apply additional cuts to minimize overlap between jets which will be discussed later. We estimate these samples to be about 95\% \bbbar, with the remaining 5\% coming from charm contamination, which is accounted for as a small systematic uncertainty.


Since jets from \bbbar\ events tend to have a much lower transverse energy than those from \ttbar\ events, the possibility that a different calibration is needed for higher energy jets must be taken into account. To this end, we bin our \bbbar\ jets according to their energy, and derive the needed correction bin-by-bin, which we apply based on the measured energy of the tagged signal jet. Great care must be taken in doing this, however, as this kinematic based correction directly introduces a jet energy scale uncertainty to the analysis. If, for example, the simulation underestimates the energies of jets, then the average decay length in a given energy bin will be too high in the simulation, throwing off the calibration. This is an unavoidable uncertainty for any kind of calibration using dijets. In fact, even if we could convince ourselves that it was unnecessary to parameterize this decay length calibration as a function of energy, a significant jet energy uncertainty would still be needed to cover the determination of which jets pass selection in the simulation. As our goal is to minimize the calorimeter jet energy uncertainties, we choose to parameterize our calibration based upon the energies of jets measured using tracking rather than the calorimetry.

\subsection{Track-based jet energies \label{sec:track_en}}

We develop a straightforward algorithm for computing the track-based energy of a jet. We start with one of our jets that has been clustered in the calorimeter in the usual manner. We select tracks that are within $R=0.4$ of the calorimeter jet direction in $\eta-\phi$ space. The tracks themselves are required to be well resolved in both the wire tracking chamber and in the silicon, and they must have a transverse momentum of at least $1\ \textnormal{GeV}/c$. We also require that the tracks pass within 3 cm along the beam axis of the fitted primary collision vertex. This cut eliminates about 85\% of the contamination from multiple interactions. We then take the total transverse component of the sum of our track four-vectors as our tracking transverse energy. We make no attempt to correct for the missing neutral particles. Our goal is not to measure the true jet energy, but rather to measure a quantity that is proportional to the true energy (in this case, the charged transverse energy) in a manner that agrees well between data and simulation. 

\begin{figure*}
\centerline{
  \subfigure[] {\includegraphics[height=2.5in]{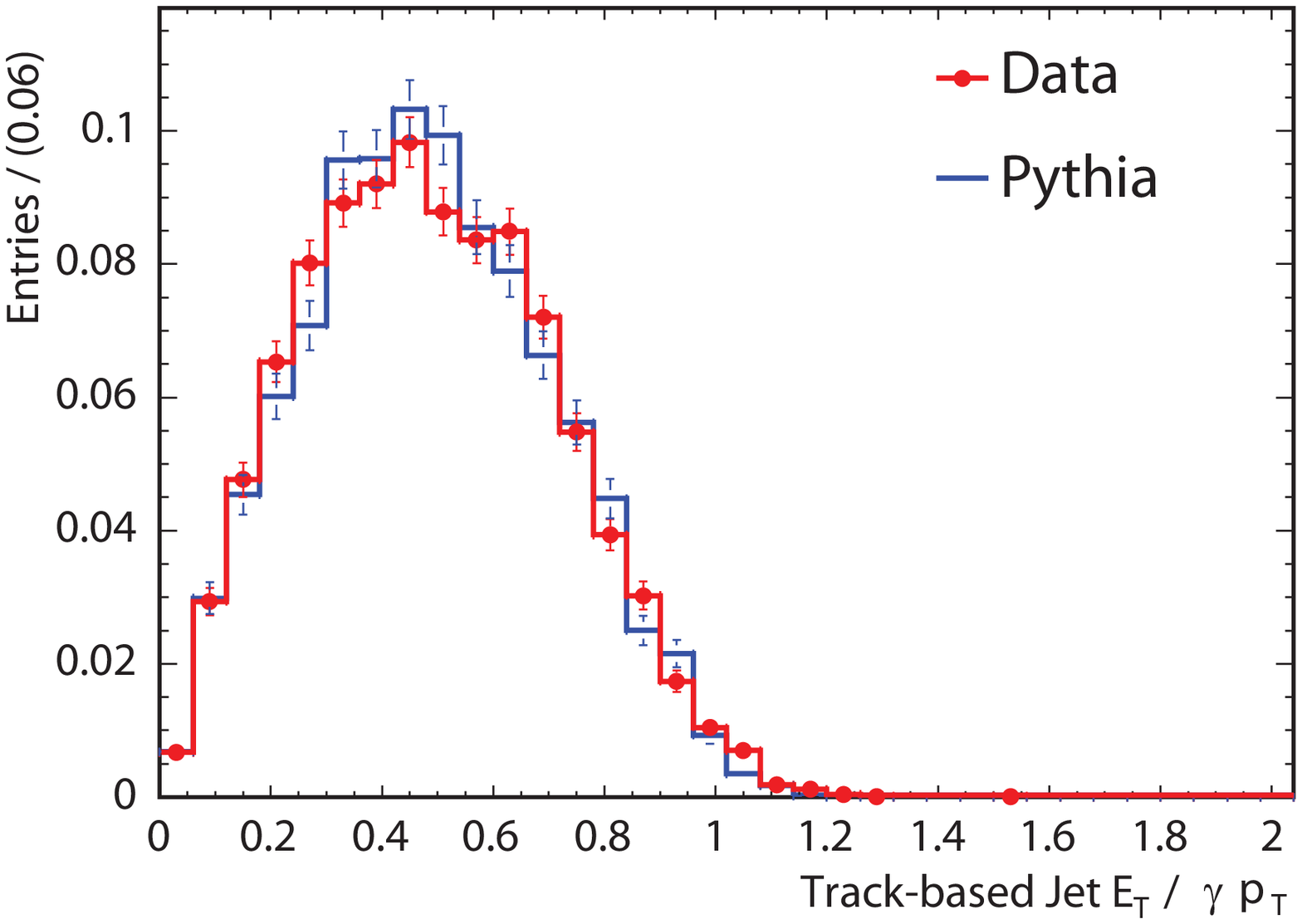} \label{fig:trk_fracs}}
  \subfigure[] {\includegraphics[height=2.5in]{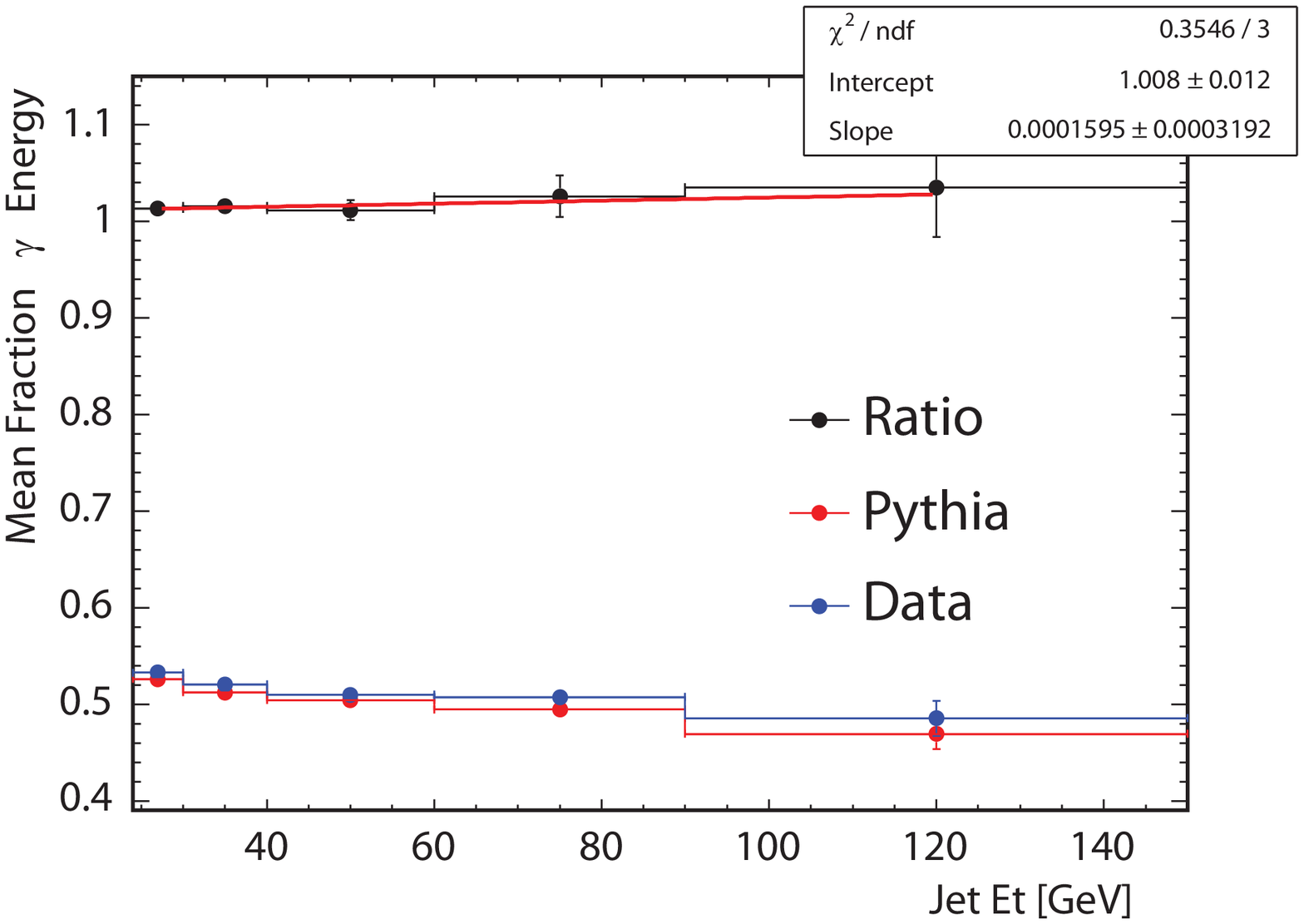}\label{fig:trk_means}}
}
\caption{(a): The distributions of track-based jet transverse energy divided by the photon $p_T$\ for events with photon $p_T$\ between $30\ \textnormal{GeV}/c$ and 40\ $\textnormal{GeV}/c$. (b): The mean fraction of track-based jet transverse energy / photon $p_T$\ for Pythia (red) and the data (blue) as a function of the measured photon transverse momentum (assumed to be the true jet transverse energy). The ratio of these trends between data and simulation is fitted to a line which is then used to correct measured track jet energies in the simulation.\label{fig:track_energy_SF}}
\end{figure*}

We calibrate the energies of our track jets with a similar approach to the one used for calorimeter jets at CDF \cite{bib:newer_CDF_JES_note}. We select events where one photon and one jet are found back-to-back ($\Delta\phi > 3$\ radians), where no extra jets in the event above 3 GeV in $E_T$ are allowed, and strict cuts on photon quality are applied to minimize fakes. Under such a selection, the transverse momentum of the photon should reflect the true transverse momentum of the jet. As a first step we consider the ratio of the track-based transverse energy of the jet to the measured transverse energy of the photon. The distribution of this ratio shows good agreement between data and simulation as seen in Figure~\ref{fig:trk_fracs} for a particular range of photon energies, and this agreement also holds for higher energy photons. The mean fraction of the photon transverse energy that is carried in the track-based jet transverse energy is shown for a range of true jet transverse energies (determined by the energy of the photon) in Figure~\ref{fig:trk_means}. The extent of the agreement between data and simulation is given by the ratio of these trends and is shown in black in the same figure. A line is fitted to these ratios and represents the calibration that will be applied to the track jet energies that we measure in our simulation. Specifically, for a given tagged jet in our simulated \bbbar\ or \ttbar\ sample, we begin by measuring the corrected transverse energy of the jet in the calorimeter and its associated track-based jet transverse energy. We then correct this track-based transverse energy according to its calorimeter-based transverse energy and the fitted function in Figure~\ref{fig:trk_means}. Uncertainties in the simulation of the calorimeter-based jet energy measurements are found to have a negligibly small impact on the correction factor that is determined in this manner. There are, however, significant statistical uncertainties on the fitted function that translate into a systematic uncertainty on our results as will be explained in Section~\ref{sec:systs}.

When applying this calibration to the simulated \bbbar\ and \ttbar\ samples, however, it is important to account for the fact that these events contain more jets than those in the photon calibration sample, and tracks from other jets may fall into the jet cones and bias the measured track jet energies upwards. If our decay length calibration parameterization proves to have a trend over jet energy, then such biases in the \ttbar\ sample will have a direct impact on the correction factor that is applied to a given jet. On the other hand, in the \bbbar\ samples there will only be a bias to the extent that \textsc{pythia} does not properly model the amount of jet overlap that is observed in data. We start by minimizing this problem as much as possible by removing events where a second jet is close to the jet we are studying and might contribute overlapping particles in both data and simulation. For our \bbbar\ events we veto events where either of our tagged jets is between $\Delta R$ of 0.7 and 1.2 of any other jet with energy greater than 9 GeV. This cut was chosen to remove most events that might have extra jets in a region that could overlap our primary jet without eliminating jets with energetic out-of-cone QCD radiation. 

\begin{figure*}
\centerline{
  \subfigure[] {\includegraphics[height=2.5in]{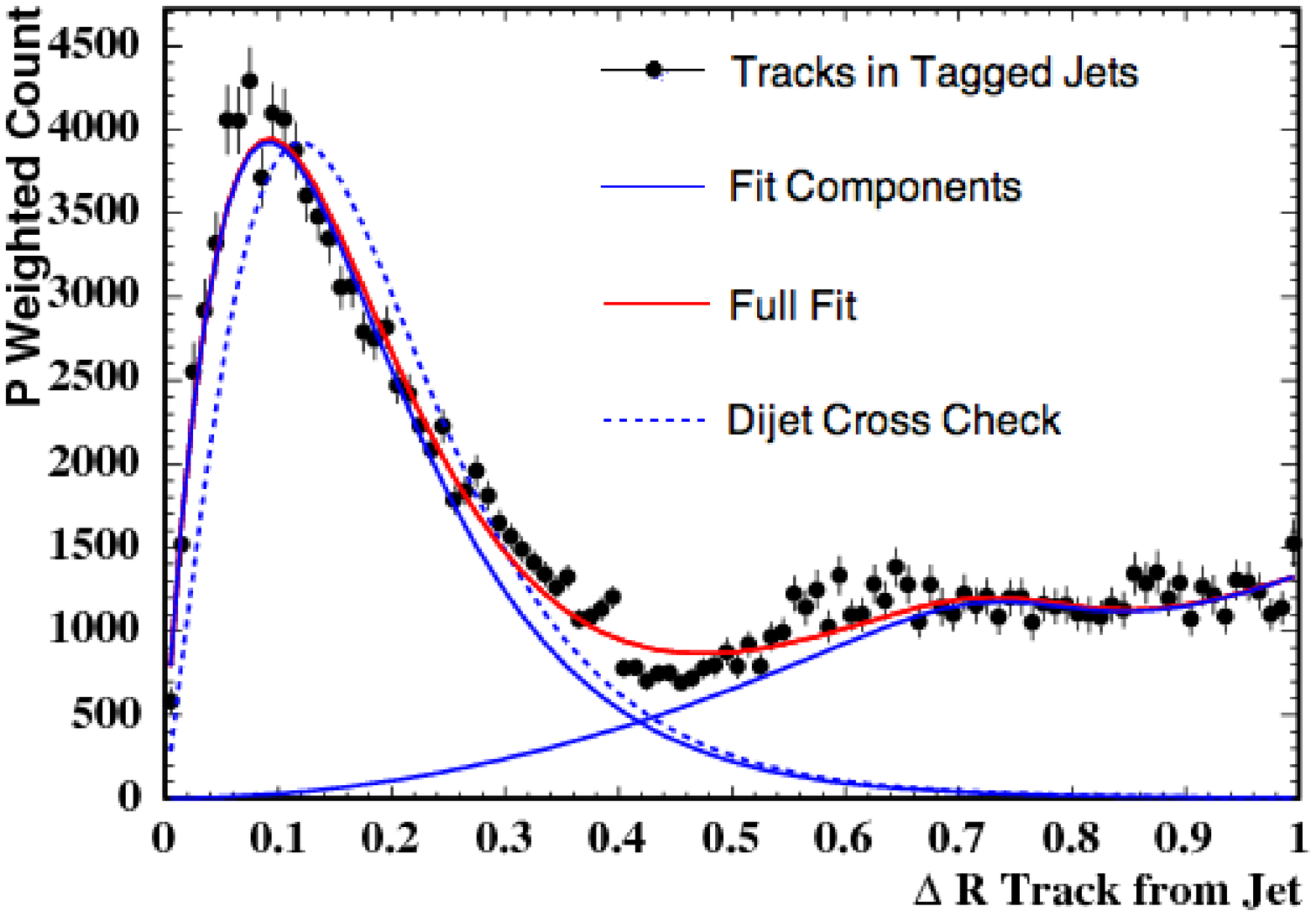}}
  \subfigure[] {\includegraphics[height=2.5in]{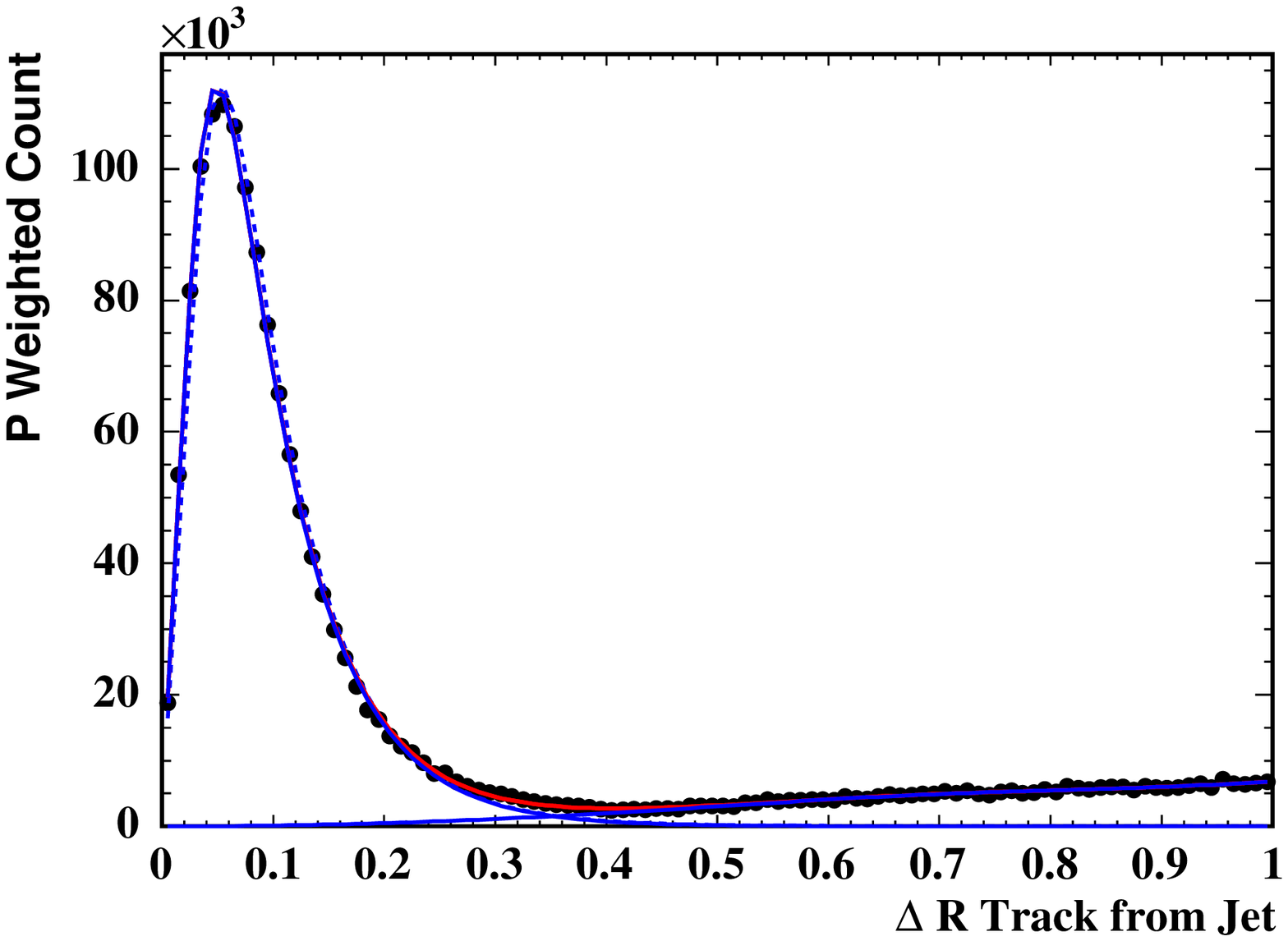}}
}
\caption{These plots show the charged jet momentum spread in $\Delta R$ for \textsc{pythia} \ttbar\ events. Specifically, the $\Delta R$ of each track relative to the calorimeter jet direction is plotted as the black points for tagged \ttbar\ jets, weighted by the track momentum. The results are then fit to determine the fractions of primary jet energy and overlap energy from other jets for \ttbar. The solid blue curves are the fit shapes for the primary and overlap fit components and the solid red curve is the full fit result. The dashed blue curve shows the expected primary distribution based upon fits in the \bbbar\ sample. (a): results for \Bot-hadron Pt less than $20\ \textnormal{GeV}/c$. This is the only kinematic range where there is any significant disagreement between the \ttbar\ and \bbbar\ results. Much more typical is (b): results for \Bot-hadron Pt between $50\ \textnormal{GeV}/c$ and $60\ \textnormal{GeV}/c$.\label{fig:ooc_cor}}
\end{figure*}

In the \ttbar\ simulation, we do not veto events in this manner. Instead, we develop a correction procedure to remove the effects of tracks from other jets that overlap our jet cones. We do this by plotting the track momentum associated with \Bot-tagged \ttbar\ jets as a function of the $\Delta R$\ between the track and the jet direction as measured in the calorimeter. We plot distributions binned in the generator-level $p_T$\ for the parent \Bot-hadron in our jet. The higher the energy of the \Bot-hadron, the narrower the $\Delta R$\ distribution, as expected. We fit these distributions to two components, a ``primary" part, and an ``overlap" part. The tracks originating from the \Bot-hadron and associated fragmentation products (the primary jet) are modeled by a Gaussian multiplied by a Fermi function to force the momentum fraction to converge to zero at small $\Delta R$. The contribution from underlying event, minimum bias, and other jets (overlap), turns out to be well modeled by a quadratic function in $\Delta R$, multiplied by a Fermi function to account for damping effects as tracks pass out of the active detector range. The fits are performed separately for each region of \Bot-hadron $p_T$. Examples of these fits (along with cross-check fits for \bbbar\ events) are shown in Figure~\ref{fig:ooc_cor}. We then use these fit results to extract correction factors, parameterized by the \Bot-hadron's $p_T$, to remove the average amount of momentum from charged particles expected to fall inside the track jet cone from other sources. 

There are a number of approximations and assumptions that have gone into these vetoes and corrections. For example, the fitting functions could be inappropriate, or the vetoes applied to the \bbbar\ events might be inadequate. If so, then using an alternate cone size to select track jets will lead to a direct systematic shift. Thus, the systematics for this procedure are evaluated by changing the track jet cone size to 0.7, removing overlap according to the revised jet size, and repeating the analysis, as explained in Section~\ref{sec:Lxy_uncertainties}. However, two other cross-checks are also performed to improve confidence in the procedure. First, we have no physical motivation for using a quadratic function as the base shape of our overlap, so in one cross-check we repeat the analysis, modeling our overlap as though it were distributed perfectly uniformly in $\eta-\phi$\ space (a line times a Fermi function in $\Delta R$). There is, of course, no physical basis for using this symmetric distribution either, since complicated correlations between jets according to \ttbar\ kinematics and sculpting from the jet clustering algorithm could render the shape asymmetric. This alternate shape is simply a cross-check that leads to an overestimated amount of overlap falling inside of the jet cone. When the analysis is repeated with the alternate shape it leads to a top-quark mass measurement result that is shifted by an amount that is smaller than the systematic we will eventually assign for the overlap removal procedure. As a second cross-check we look at the tagged jets that are selected to be back-to-back with the muon jet in the \bbbar\ sample. If the overlap fitting function is incorrect, then we would expect the shape of the primary part of the fit to be better modeled in these \bbbar\ events, since the overlap is about a factor of four smaller than in \ttbar\ events. These \bbbar\ comparison distributions are shown as the dashed curves in Figure~\ref{fig:ooc_cor}. By far the worst agreement is seen in the lowest bin of hadron $p_T$ (shown on the left). In every other bin of \Bot-hadron momentum, the differences between the \bbbar\ and \ttbar\ results can barely be distinguished, as is the case in the right hand plot. But since less than 8\% of \ttbar\ jets have hadron momenta that fall into the lowest bin, the systematic mass shifts caused by enforcing the alternate \bbbar\ shapes in the primary distribution will be well less than the systematic that we will end up claiming.

\subsection{Final decay length calibration \label{sec:final_lxy_calib}}

\begin{figure}  \begin{center}
\includegraphics[height=2.5in]{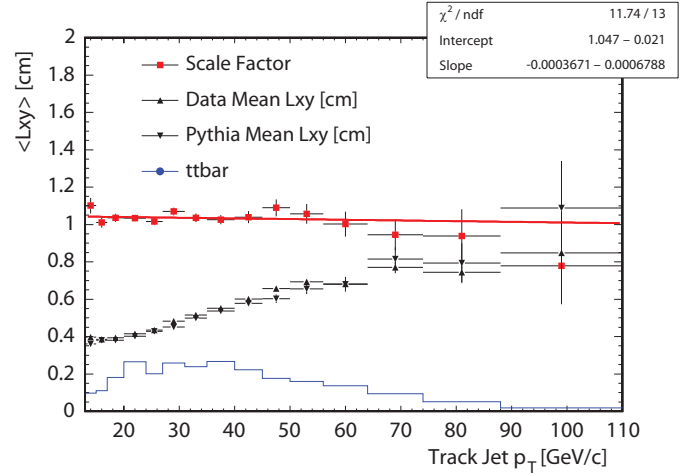}
\caption{The decay length calibration parameterization versus track jet transverse energy is shown here. It is developed from the solid black data and simulated average decay lengths, the ratio of which gives the Scale Factor points, which are then fit to determine the calibration. Also shown is the distribution of \ttbar\ jets to which the calibration is applied. For the Lxy data and simulation the vertical scale represents the mean decay length in cm, while for their Scale Factor ratio and the \ttbar\ distribution this axis is unitless.\label{fig:SF_param}}
\end{center}\end{figure}

With the track jets in hand we can evaluate the decay length calibration parameterization in the \bbbar\ sample as described above. The calorimeter is still used to select jets and determine their direction, but we loosen the calorimeter jet energy cut significantly, and exclude most jets with energies near this cut by applying a track jet transverse energy cut. As a result, calorimeter driven jet energy uncertainties will play a minimal role. Additionally, we apply the overlap vetoes and corrections as described above. Since it is possible for a jet to pass our tracking energy cuts while failing the calorimeter energy selection, fluctuations within standard, calorimeter-based jet energy uncertainties can still cause events to pass in and out of selection. However this is a small effect which is only significant in the lowest energy track jet bins, and thus leads to a small systematic uncertainty.  Figure \ref{fig:SF_param} shows the trends in the mean decay length for data and simulation. The ratio of these trends gives us the correction that should be applied to the measured decay lengths in data, depending on the measured track jet energy in the signal samples.  The distribution of the track jet energies to which the calibration will be applied is overlaid.

\section{\label{sect:method} Mass Measurement Method \label{sec:method}}

Given our mean measured Lxy and lepton \pT\ values from the data (associated with the distributions of Figure~\ref{fig:meas_stacks}) we need to determine the associated top-quark masses and statistical uncertainties. We simulate experiments under a variety of hypothetical top-quark masses and use them to perform each of the measurements as described below.

\subsection{Single variable measurements}

Pseudoexperiment events are drawn from the events that are used to construct the signal and background distributions where the probability of each event is given by its PDF weighting as discussed in Section \ref{sec:corrections}. The mean Lxy (\pT) of the tagged jets (leptons) in these events will be used to measure the mass. Samples of \ttbar\ events are generated under 23 hypothetical top-quark mass values ranging from $140\ \textnormal{GeV}/c^2$ to $220\ \textnormal{GeV}/c^2$, and the decay length results are corrected according to the track jet energies by the fit results of Figure~\ref{fig:SF_param}. 

Uncertainties from the background normalization are small and are wrapped into the pseudoexperiments. A total of $92.5 \pm 17.1$ background events are expected, however due to the iterative nature of the \ttbar\ cross section evaluation, there is a small additional uncertainty on the background normalization of $\pm 3.7$\ events based upon the theoretical uncertainty on the input \ttbar\ cross section, leading to a total background count uncertainty of $\pm 17.5$. For each pseudoexperiment, the total number of background events is fluctuated according to a Gaussian with the above mean and RMS to determine an expected background normalization. The resulting number is then fluctuated according to Poisson statistics to determine a background normalization for each pseudoexperiment. Given a fixed number of observed data events, the excess is taken to be signal. Further, for each pseudoexperiment the Lxy calibration parameterization is fluctuated within the fitted statistical uncertainties shown in Figure~\ref{fig:SF_param}.

\begin{figure*}
\centerline{
  \subfigure[] {\includegraphics[height=2.5in]{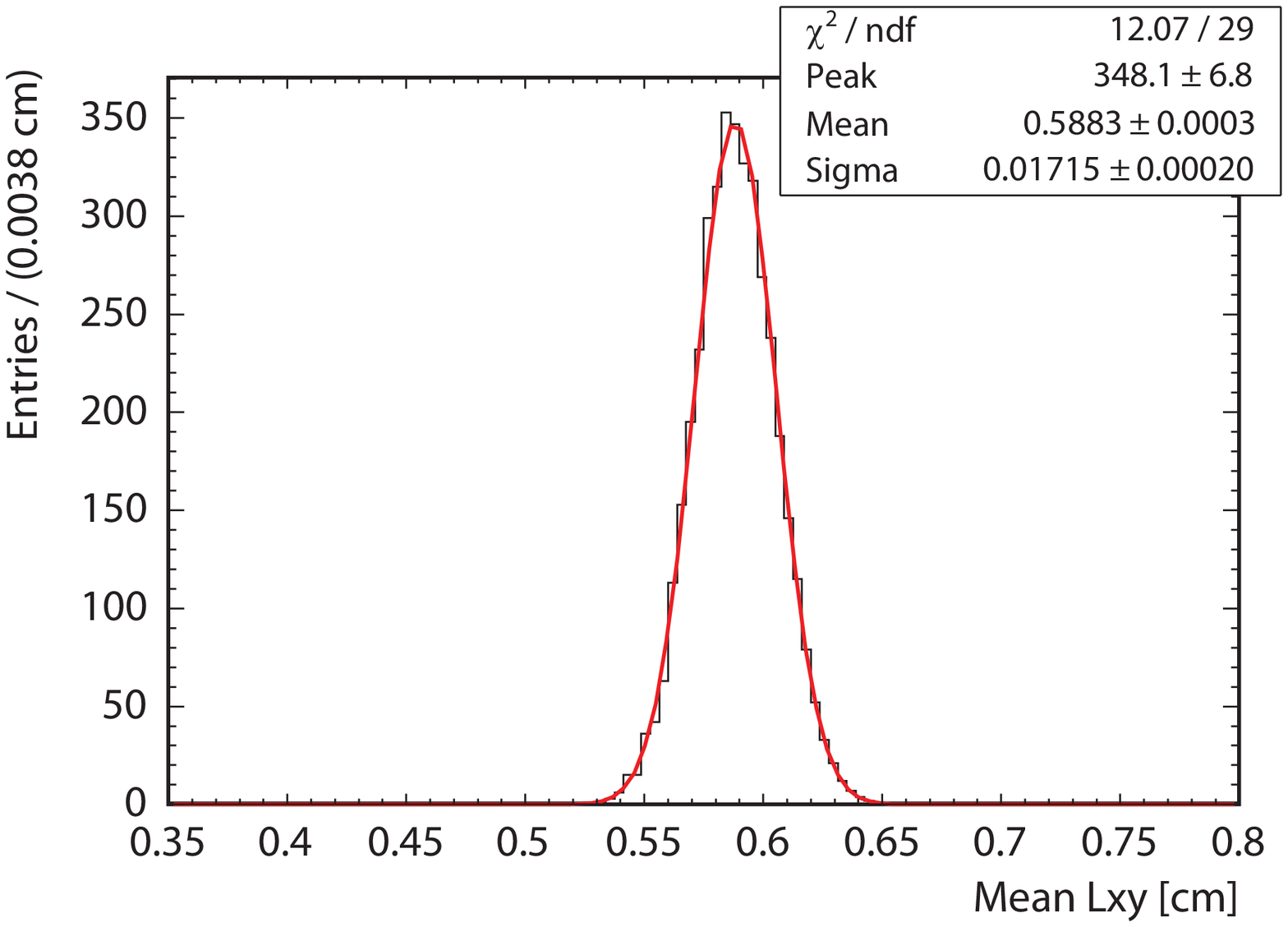}}
  \subfigure[] {\includegraphics[height=2.5in]{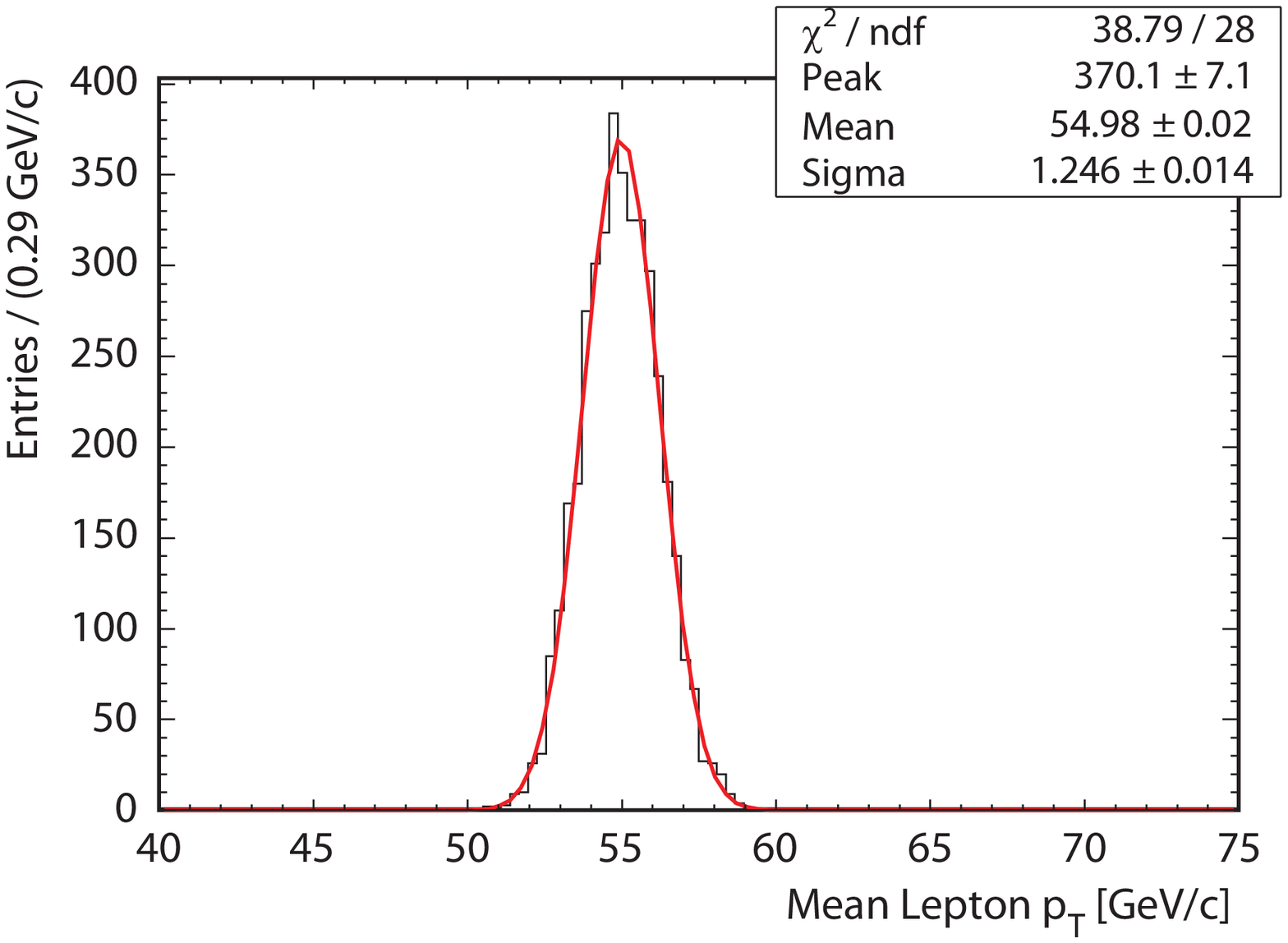}}
}
\caption{Pseudoexperiment distributions and fit results for example hypothetical top-quark mass values. (a): Mean Lxy pseudoexperiments for $m_t = 165$\ GeV/$c^2$. (b): Mean lepton $p_T$\ pseudoexperiments for $m_t = 173$\ GeV/$c^2$. The results of these fits will be used to construct Figure~\ref{fig:oneD_results}.\label{fig:example_pe_res}}
\end{figure*}

\begin{figure*}
\centerline{
  \subfigure[] {\includegraphics[height=2.5in]{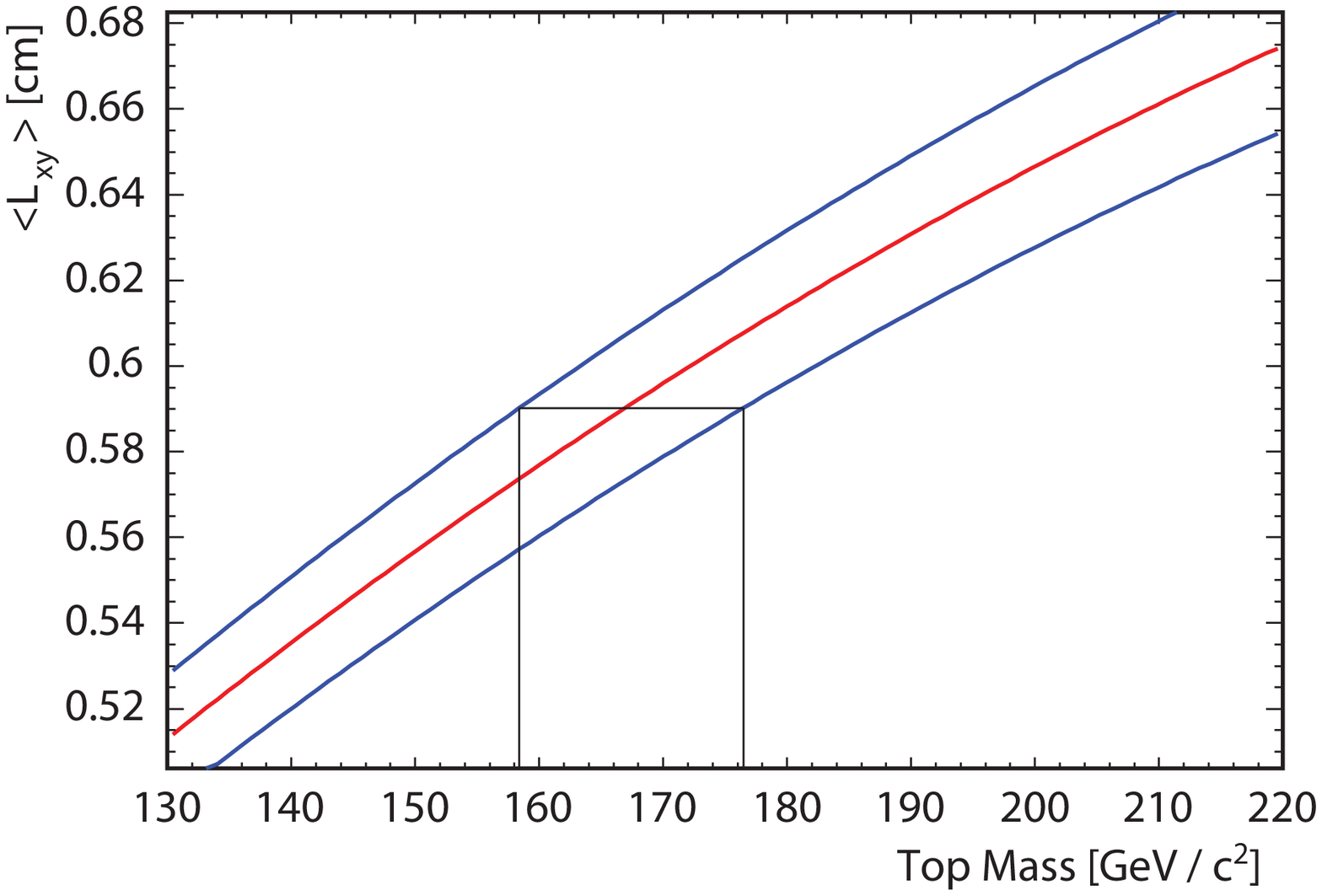}}
  \subfigure[] {\includegraphics[height=2.5in]{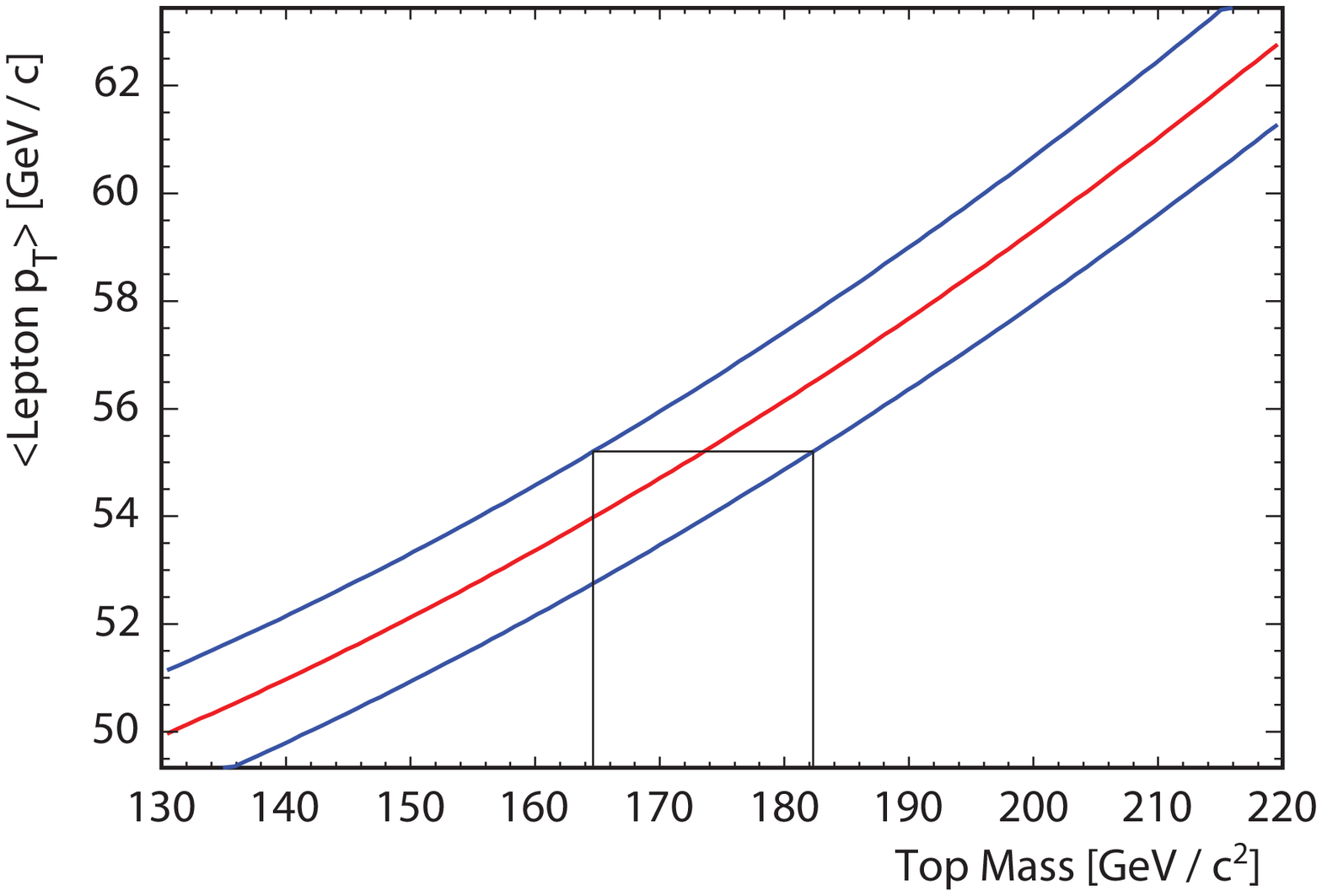}}
}
\caption{The mean values of the Gaussian fits to the pseudoexperiment results from Figure~\ref{fig:example_pe_res} are fit to the quadratic polynomials plotted here in red. The one sigma statistical uncertainties from these fits lead to the blue contours. The mean values from data are shown as the horizontal black lines. These values are then translated into top-quark masses according to their intersections with the red polynomial, and into asymmetric statistical uncertainties according to their intersections with the blue polynomials.}\label{fig:oneD_results}
\end{figure*}

The mean Lxy and lepton $p_T$ pseudoexperiment results are Gaussian in shape and are fit to a Gaussian for each hypothetical top-quark mass. Examples of these fits are shown in Figure~\ref{fig:example_pe_res}. To evaluate the top-quark mass results for the Lxy and lepton $p_T$ measurements, the central values of these Gaussians are plotted as a function of top-quark mass and are fit to a quadratic polynomial. The mean Lxy and lepton $p_T$ values measured in data are then converted to the measured top-quark mass values according to this polynomial. To extract the statistical uncertainties, the central Lxy and lepton $p_T$ pseudoexperiment values are shifted up and down by the standard deviation of these Gaussian fits. Then the difference between the measured top-quark mass according to the unshifted polynomial and the top-quark masses resulting from these one standard deviation shifts are taken to be the asymmetric one sigma statistical uncertainties on the measurements. The fitted polynomials are shown in Figure~\ref{fig:oneD_results}. The mean Lxy and lepton $p_T$ values measured in data are also shown as the horizontal black lines, along with the projections that are used to determine the statistical uncertainties. 


%

\subsection{Measurement using both variables}

\begin{figure*}
\centerline{
  \subfigure[] {\includegraphics[height=2.5in]{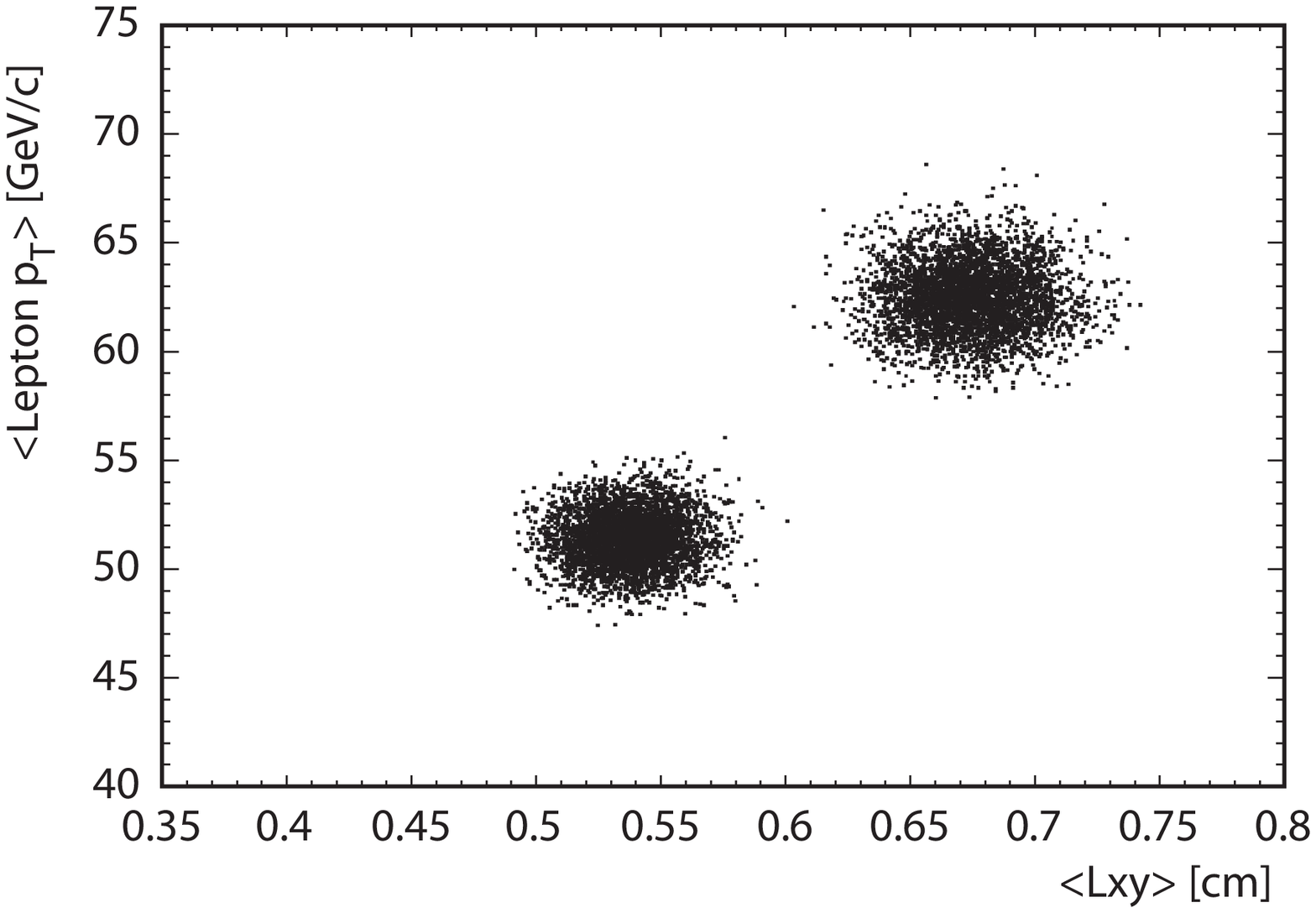} \label{fig:PE_scat}}
  \subfigure[] {\includegraphics[height=2.5in]{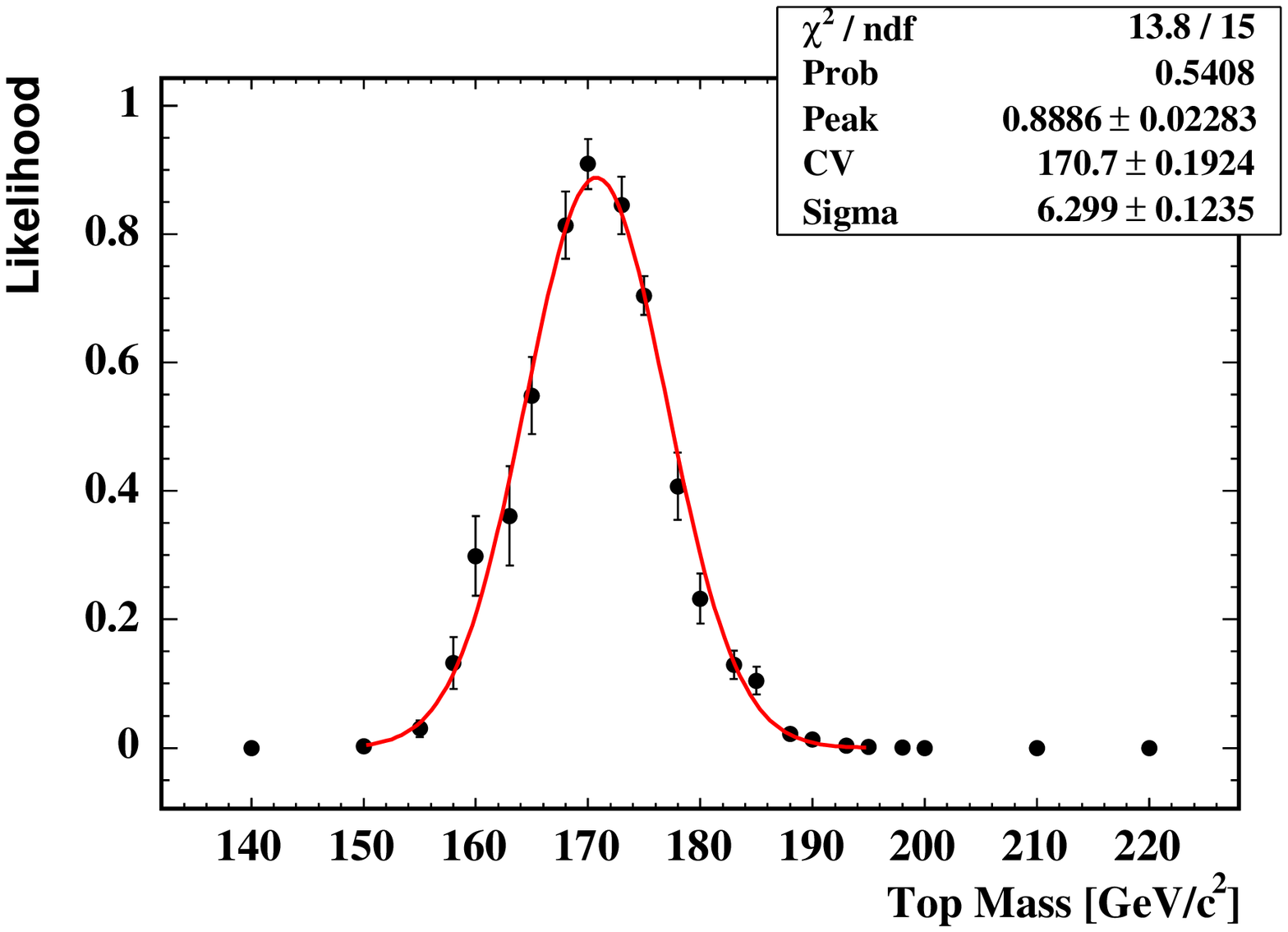} \label{fig:twod_like}}
}
\caption{(a): distribution of mean Lxy versus mean lepton $p_T$ from pseudoexperiments for extreme mass cases of 140 GeV and 220 GeV. (b): the pseudoexperiment results for the 23 mass points considered are used to determine the likelihood of agreement with the data according to the metric of Equation \ref{CorrEquation}, and the results are plotted and fitted here. The mean fit result is taken as the measurement result, and the RMS represents our statistical uncertainty.\label{fig:2d_means_and_data_fit}}
\end{figure*}

The pseudoexperiments from the single variable results are used to plot two-dimensional mean Lxy versus mean lepton $p_T$ distributions. 
The results for the two most extreme mass hypotheses are overlaid in Figure~\ref{fig:PE_scat}. 
                                                                                                     
The observed data produce a point on this two-dimensional plane. Given this point, our task is to determine the most likely value of the top-quark mass and the associated statistical errors. To accomplish this, we evaluate a likelihood for each
mass hypothesis according to the data. The likelihood is simply the probability
that if the true mass were the one in our hypothesis, the mean Lxy and
lepton $p_T$ results would fluctuate as far away or farther than the results we see in
the data point. This probability is taken from pseudoexperiment results such as
those shown in Figure~\ref{fig:PE_scat}. Specifically, we
evaluate a ``distance" that our data point is from the expected central value
for our mass hypothesis, and take the likelihood that the hypothetical mass is
correct to be the fraction of pseudoexperiments which
are ``farther away" from the expected values than our data point.

For this approach to be meaningful, a reasonable definition of
``distance" must be used. We choose our definition so that distances are
equal for points along the equal probability contours of a two-dimensional
Gaussian centered at the expected mean Lxy and lepton $p_T$ values, with the expected
standard deviations. Here, the expected means and standard deviations are taken from the Gaussian fits to the mean Lxy and lepton $p_T$\ pseudoexperiment results that were described for the single variable measurements. Then, the ``distance" from
the expected value is defined according to the metric in Equation \ref{CorrEquation}:

\begin{equation}
D = \sqrt{ \left(\frac{\delta P_t}{{{\sigma}_P}_t}\right)^2 + \left(\frac{\delta L_{xy}}{{{\sigma}_L}_{xy}}\right)^2 } 
\label{CorrEquation}
\end{equation}
                                                                                                     
Here, $\delta P_t$ ($\delta L_{xy}$) is the difference between the mean lepton $p_T$ (Lxy) of the data and the
fitted central value for the hypothesized top-quark mass, and ${{\sigma}_P}_t$ (${{\sigma}_L}_{xy}$) is the fitted standard deviation of the mean lepton $p_T$  (Lxy) for the hypothesized top-quark mass. In principle this approach could be modified to account for a
correlation between the two variables, but this is unnecessary as the correlations are empirically determined to be negligibly small.                                                                                         
                                                                                                     
Based on this equation, the fraction of pseudoexperiments for each hypothesis for which the
distance metric is evaluated to be larger than that for the data is taken as the
likelihood for the hypothesized mass. Finally, the likelihood values
for each mass point are plotted with statistical uncertainties determined by
adding in quadrature the statistical uncertainties from the pseudoexperiments, and the statistical uncertainty due to the number of simulated events from which they are drawn. We run enough pseudoexperiments (4000) that the size of our simulated \ttbar\ samples is the primary limitation. These likelihoods are fit to
a Gaussian as shown in Figure~\ref{fig:twod_like}. The mean of the Gaussian is taken to be the result of our combined measurement with a statistical uncertainty given by the RMS of the fit.

\subsection{Results}
Data events passing event selection have a mean Lxy of \dataLxy\ and a mean lepton \pT\ of \dataLepPt. Based upon these values, the mass measurement with the decay length technique yields a result of \lxyStatRes, and with the lepton transverse momentum technique yields a result of \lepPtStatRes, where the errors are statistical only. For the simultaneous measurement with both variables the fit to data is shown in Figure~\ref{fig:2d_means_and_data_fit} and corresponds to a mass result of \combStatRes. 

Some sanity checks were run for each of the three top-quark mass measurements. For these checks, nineteen additional top-quark mass samples were generated with top-quark masses varying between 152 and 193 $\textnormal{GeV}/c^2$. Ten of these samples were blind (the masses were hidden from the authors until after the measurements were finished). Pseudoexperiments were thrown using these samples, and the means of the measured top-quark mass results proved to be consistent with expectations to within the statistical uncertainties, indicating the method is unbiased. Further, the pulls (defined as the difference between each measured mass and the generated mass divided by the statistical uncertainty) of the pseudoexperiment results prove to have a width that is consistent with 1.0, indicating that the estimated statistical uncertainties are reliable.


\section{Systematic uncertainties \label{sec:systs}}

Next we will discuss the systematic uncertainties for this measurement, and the motivations for the procedures we apply. A list of our final systematic uncertainties can be found in Table~\ref{final_syst_table}.

\subsection{Background uncertainty}
As mentioned above, the uncertainty on the background composition and shape is evaluated in the control regions of the one- and two-jet bins, shown in Figures~\ref{fig:onejet_cross_check} and ~\ref{fig:twojet_cross_check}. The differences between the observed and expected means are shown in Table~\ref{bkg_cross_res_table}. The largest disagreements are observed in the one-jet bin of Lxy, and in the two-jet bin for lepton $p_T$. Since both of these worst-case shifts are larger than their uncertainties, they are taken as the uncertainties on the background mean results and are scaled by the background fraction to determine the systematic errors.

\begin{table}[th]
  \begin{center}
  \caption{Background shifts and uncertainties in the one and two-jet control regions. The uncertainties account for both statistical effects due to data limitations as well us uncertainties in the relative contributions from the individual backgrounds from the cross section measurement.}\label{bkg_cross_res_table}
  \begin{ruledtabular}
  \begin{tabular}{lcc}
Variable & Shift \\
  \hline
One-Jet Lxy & $-0.0131 \pm 0.0082 (\textnormal{cm})$\\
One-Jet lepton $p_T$ & $0.25 \pm  1.25 (\textnormal{GeV}/c)$\\
Two-Jet Lxy & $0.0022 \pm  0.0118 (\textnormal{cm})$\\
Two-Jet lepton $p_T$ & $-2.22 \pm  1.05 (\textnormal{GeV}/c)$\\
  \end{tabular}
  \end{ruledtabular}
  \end{center}
\end{table}

\subsection{QCD radiation uncertainty}

Uncertainties in the simulation of QCD radiation could have a significant impact on the results of an analysis like this which is dependent upon an accurate modeling of the boost of the decay products of the \ttbar\ system. Comparisons of the dilepton boost for Drell-Yan events have been made between data and simulation \cite{bib:PDFReweightRef} and used to constrain inaccuracies in the modeling of initial state radiation in quark-antiquark interactions. \textsc{pythia} parameters were varied to conservatively bracket the possible disagreement between data and simulated initial state radiation, and analogous parameters were simultaneously varied for the final state radiation by an equivalent amount. Signal events were generated with these parameters shifted up and down, and these samples were compared with each other and the nominal \textsc{pythia} sample. Half of the largest mass shifts between any pair of these three samples was taken as the QCD radiation systematic uncertainty. 

\subsection{Parton Distribution Function uncertainty}

As described in Section~\ref{sec:corrections}, for this analysis we reweight our events to match both the predictions of the next to leading order CTEQ6M parton distribution function, and the expected gluon fusion top production fractions predicted by theory. The advantage of using the CTEQ6M PDF is that it includes a prescription for estimating PDF uncertainties~\cite{bib:CTEQ6M_ref}. The degrees of freedom of the PDF can be parameterized in 20 orthogonal sources of uncertainty that are commonly called eigenvectors. For each of these uncertainty sources there are two parton distribution functions, where the parameters associated with these eigenvectors are shifted up or down to cover a 90\% confidence interval. We reweight our top-quark mass $175\ \textnormal{GeV}/c^2$ sample to each of these forty alternate PDFs, taking half the full mass shift for each pair as a systematic uncertainty, and adding them in quadrature. While these uncertainties are intended to represent a 90\% confidence interval, we conservatively use them as one sigma systematics instead. While we fix the fraction of \ttbar\ events produced by gluon fusion interactions to theoretical expectations, it should be noted that some of these eigenvector variations are expected to change this fraction. Thus, we allow the gluon fractions to float around their expectations for purposes of determining systematic uncertainties on the PDF results. 

One uncertainty that is not accounted for by these eigenvectors is the uncertainty on the strong coupling constant, $\alpha_s(m_Z) = 0.1176 \pm 0.0020$ \cite{bib:PDG_ref}. To study the effects of this uncertainty we reweight to the CTEQ6A and CTEQ6B PDFs~\cite{bib:CTEQ6AB_ref}, which are two different series of PDFs constructed with varying $\alpha_s$ values in intervals of 0.002. We average the mass shifts obtained when varying the PDFs, and arrive at an uncertainty that is roughly half as large as the eigenvector uncertainty. We observe consistency between the A and B series PDFs. We add this uncertainty in quadrature to the eigenvector uncertainty to determine our full PDF systematic. As a final cross-check to the results of the CTEQ Collaboration, we also reweight to the MRST Collaboration's NLO PDF MRST2004 \cite{bib:MRST2004_ref}. We observe agreement well within our stated eigenvector uncertainty when this result is compared with the corresponding CTEQ6A/B PDFs.

\subsection{Generator uncertainty}

In CDF top-quark mass analyses it is conventional to re-evaluate the top-quark mass using samples produced with the \textsc{herwig} 6.510 generator \cite{bib:Herwig_ref}, and take the shift from the \textsc{pythia} mass result as a systematic uncertainty. Note that many of the differences between these generators will double count our existing systematic uncertainties. The different fragmentation models between \textsc{pythia} and \textsc{herwig} will double count the decay length scale, jet energy, and QCD Radiation uncertainties. \textsc{herwig} also does not properly handle QED radiation off of leptons from the \W-boson decays, which is instead inserted with the PHOTOS \cite{bib:PHOTOS_ref} program. Differences in these approaches will double count our lepton energy scale uncertainty. The generators also have minor differences in the applied top width (and \textsc{herwig} has a sharp cutoff preventing the presence of top quarks in the high and low mass tails), and only \textsc{herwig} properly handles spin correlation between the two top quarks. Despite the double counted uncertainties, we follow the convention of other analyses for consistency by taking the difference between our \textsc{pythia} and \textsc{herwig} mass results as a generator systematic. For the Lxy and combined measurements the statistical uncertainty on our mass shift is greater than the shift itself, so we take the uncertainty on the shift as our systematic instead.

\subsection{Lepton momentum uncertainty}

\begin{figure*}
\centerline{
  \subfigure[] {\includegraphics[height=2.5in]{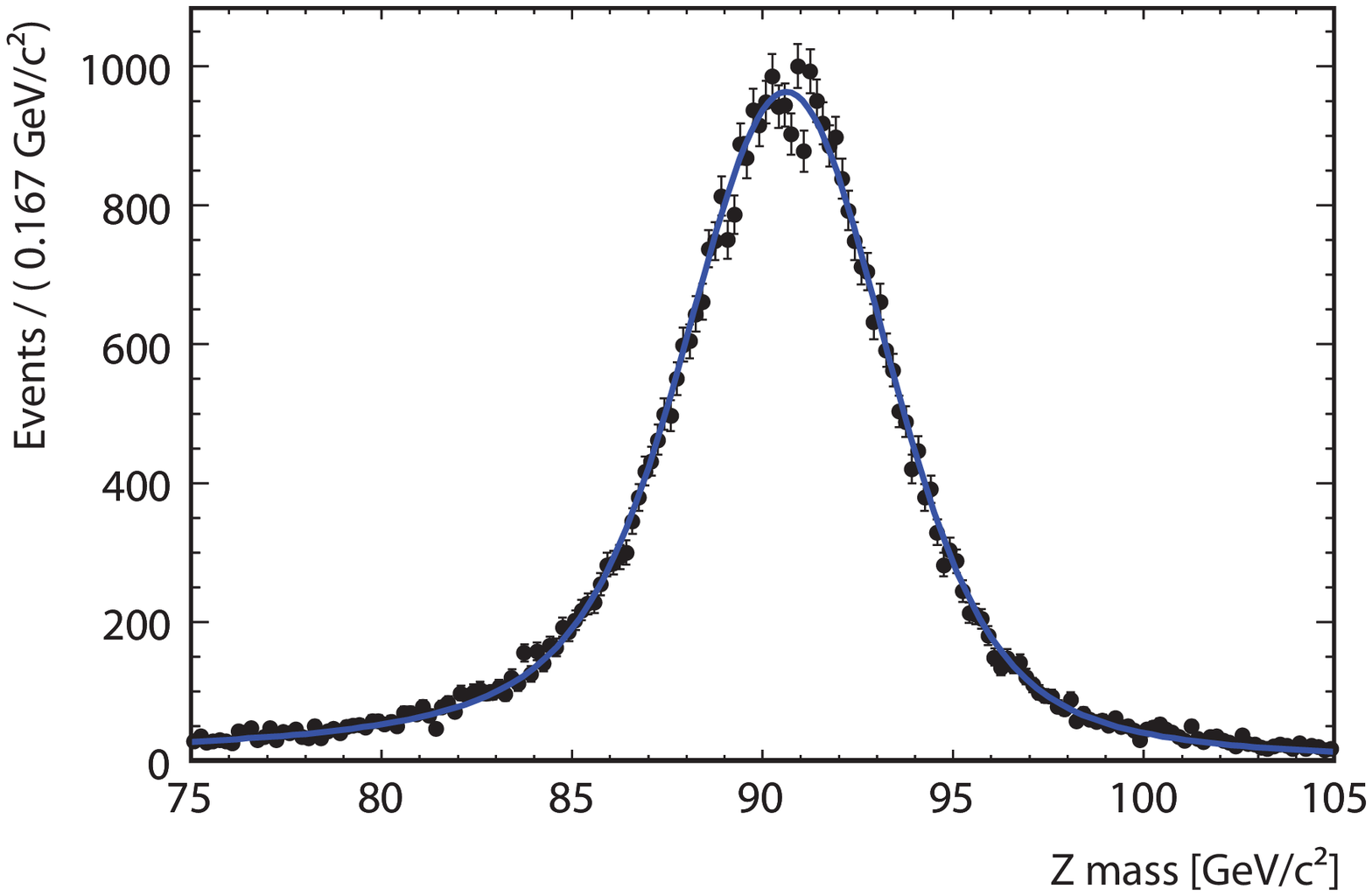}}
  \subfigure[] {\includegraphics[height=2.5in]{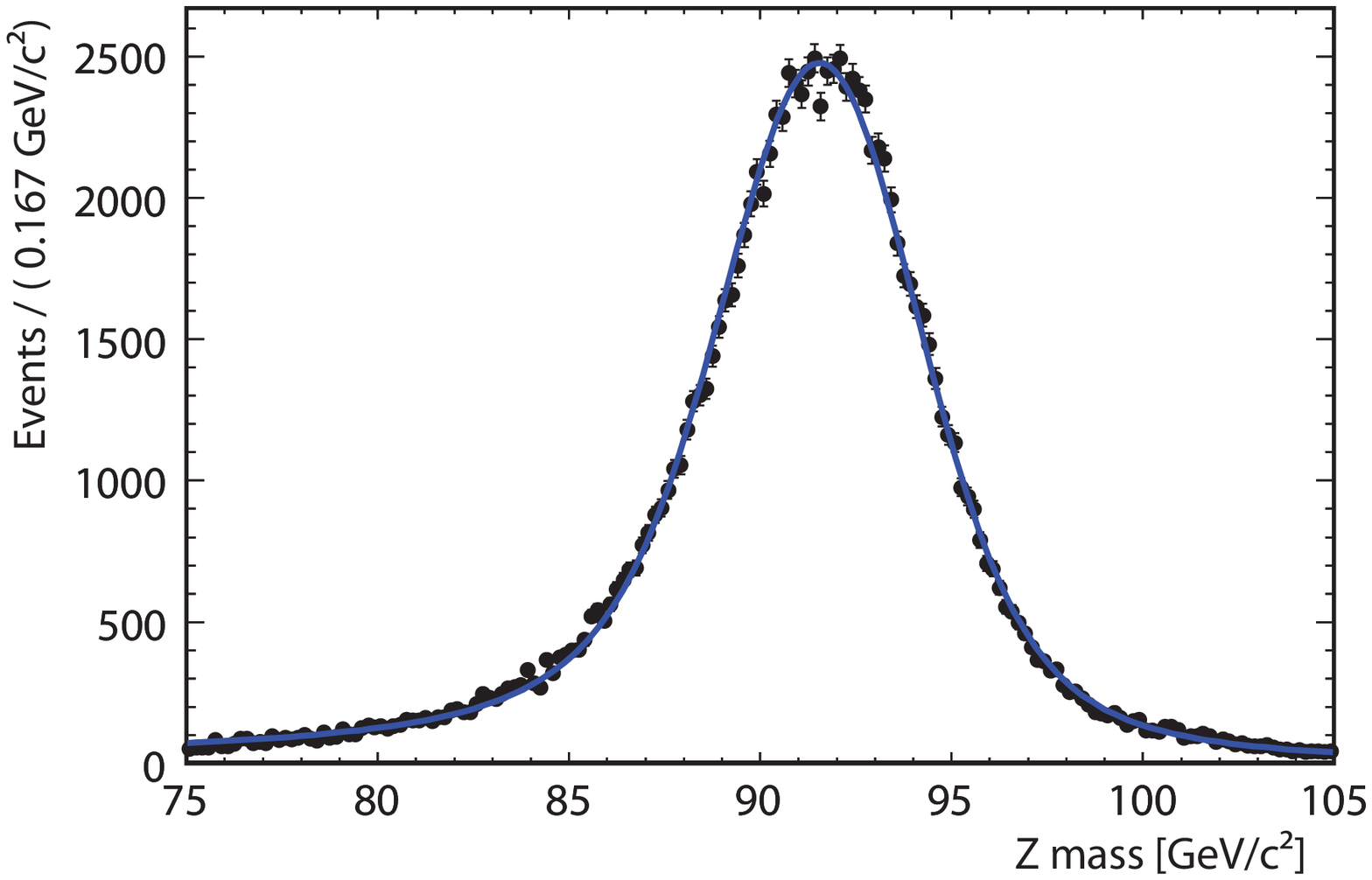}}
}
\caption{(a): fit to the dilepton mass peak for electrons in data. (b): fit to the \Z\ mass peak for electrons in simulation.\label{fig:Z_ele_fits}}
\end{figure*}

\begin{figure*}
\centerline{
  \subfigure[] {\includegraphics[height=2.5in]{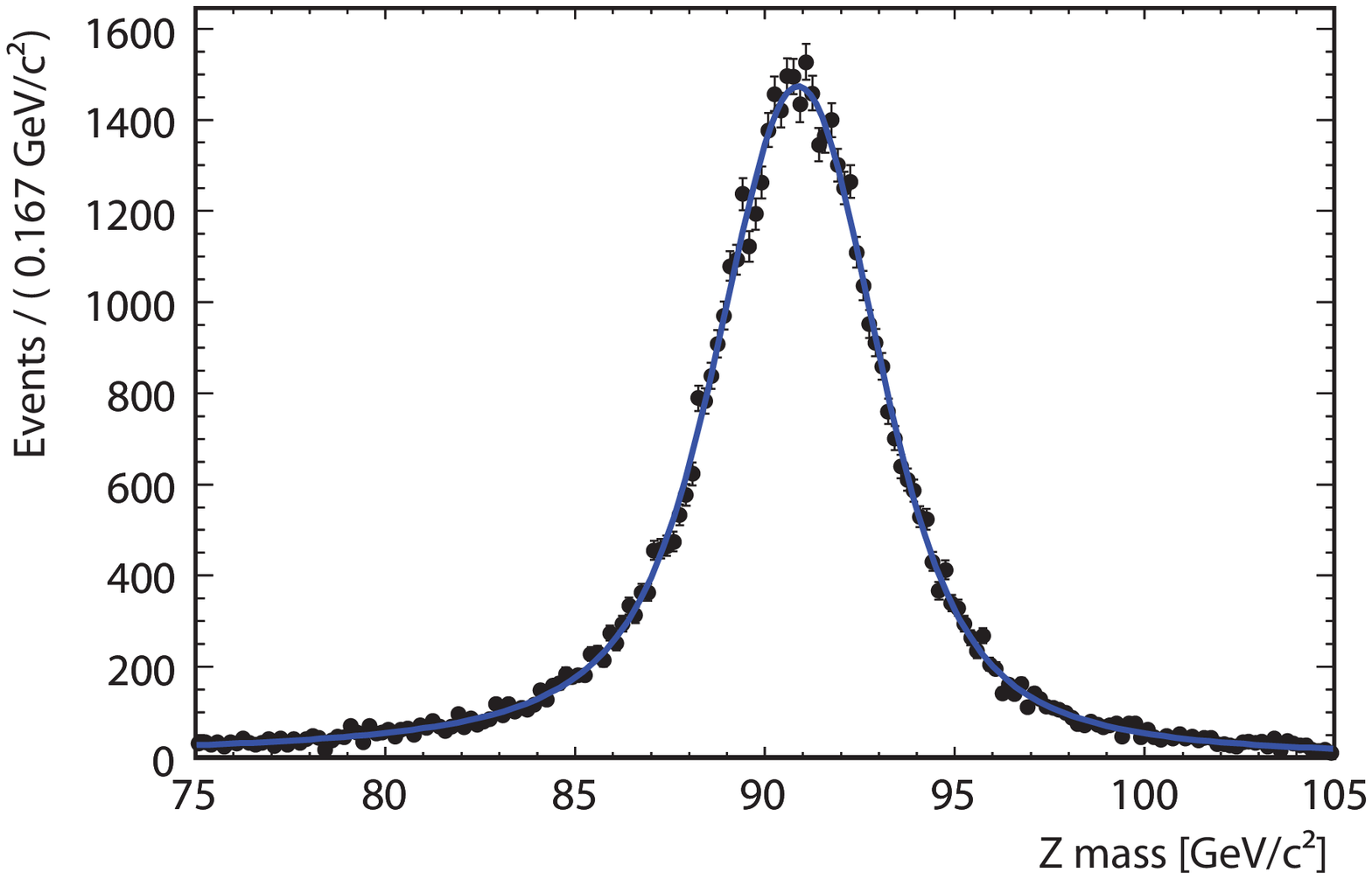}}
  \subfigure[] {\includegraphics[height=2.5in]{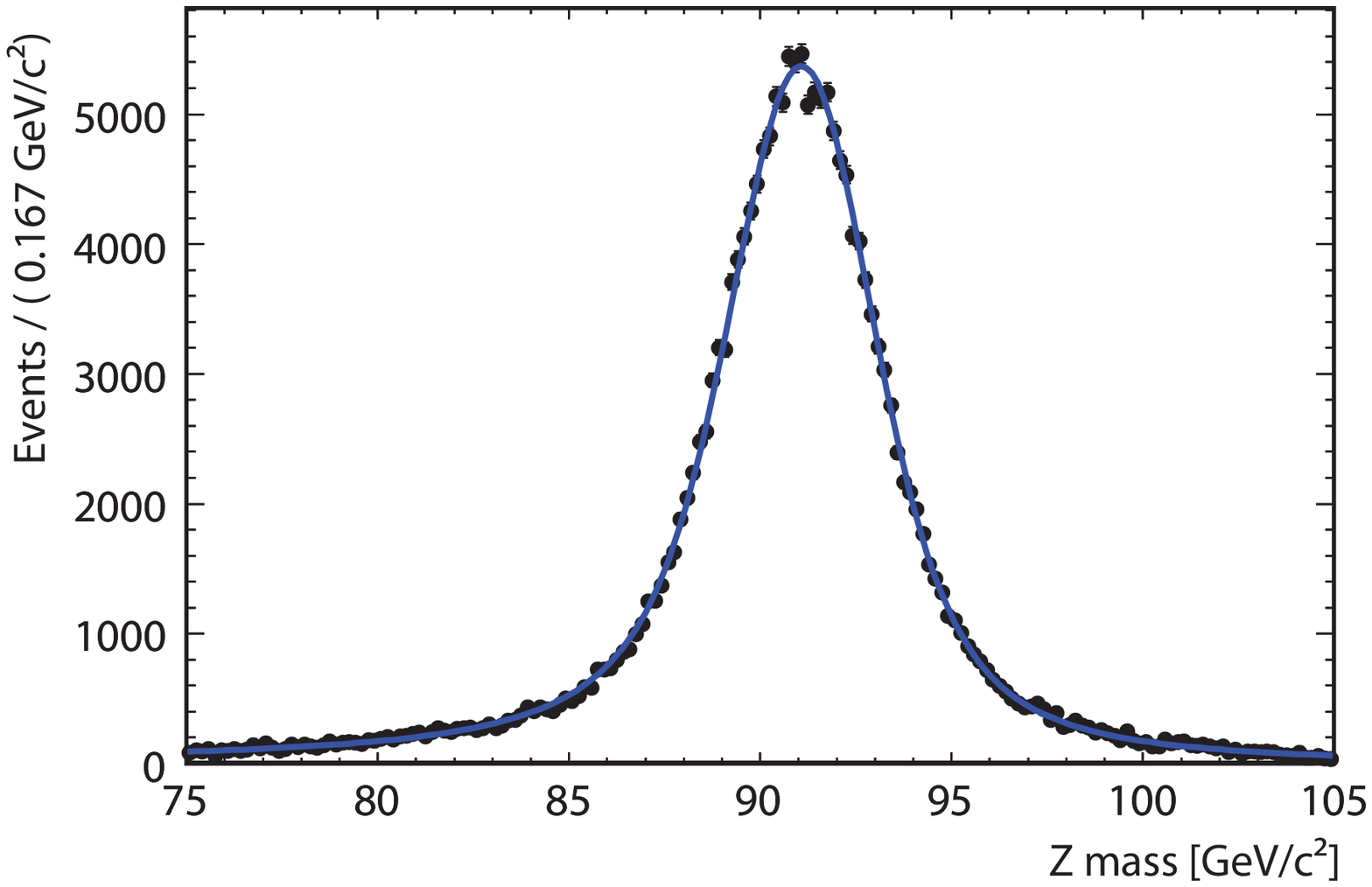}}
}
\caption{(a): fit to the dilepton mass peak for muons in data. (b): fit to the \Z\ mass peak for muons in simulation.\label{fig:Z_muon_fits}}
\end{figure*}

The modeling of the lepton momentum in simulation is tested by fitting the invariant mass of \Z's in data and simulation, separately for electrons and muons. A Breit-Wigner function is used to model the inherent \Z\ width, which was convoluted with a Gaussian to account for detector resolution. Additionally, to model the kinematic reduction in the cross section for higher mass \Z\ production this function is multiplied by a decaying exponential. When a function modeling a QCD background shape is included the fits return zero for its normalization as expected due to the high purity achieved by the lepton selection. The fit distributions are shown in Figures~\ref{fig:Z_ele_fits} and~\ref{fig:Z_muon_fits}.

The centers of the Breit-Wigner fit results, shown in Table~\ref{Z_fit_table}, were compared between simulation and data. To evaluate the systematic, the mean lepton $p_T$ of the signal was scaled by the ratio of the data and simulated means for electrons and muons separately, and the shift in the measured mass results was taken as a systematic (the statistical uncertainties on these fit results are negligible). Clearly, the disagreements in the electron results dominate this uncertainty. 


\begin{table}[th]
  \begin{center}
  \caption{Centers of the Breit-Wigner functions from the fits to the \Z\ peaks shown in Figures~\ref{fig:Z_ele_fits} and~\ref{fig:Z_muon_fits}}\label{Z_fit_table}
  \begin{ruledtabular}
  \begin{tabular}{lc}
  Sample & Mass Peak ($\textnormal{GeV}/c^2$) \\
  \hline
  Muon Simulation & 91.16 \\
  Muon Data & 90.96 \\
  Electron Simulation & 91.81 \\
  Electron Data & 90.84 \\
  \end{tabular}
  \end{ruledtabular}
  \end{center}
\end{table}

\subsection{Decay length related uncertainties \label{sec:Lxy_uncertainties}}

The procedure for calibrating the decay length measurements in our signal sample is described in section~\ref{sec:lxy_weightings}. There are many uncertainties which must be considered for this calibration, some due to the modeling of \Bot-jets, and others due to the track jet energy measurements that are used to parameterize the calibration.

The decay length calibration has a statistical limitation due to the data and \textsc{pythia} ~\bbbar\ sample sizes. This uncertainty is folded directly into the pseudoexperiments as explained in section~\ref{sect:method}, but its contribution is quite small. There are a number of uncertainties on the photon plus jet energy calibration technique. The energy scale calibration curve has an associated statistical uncertainty which propagates through to a mass uncertainty. A systematic uncertainty of 1\% is taken on the measured energy of the photon in the simulation, which corresponds to a 1\% uncertainty on the measured track jet transverse momentum. Finally, as described in \cite{bib:newer_CDF_JES_note}, about 30\% of the photon plus jets sample in data is composed of QCD dijet production where one of the QCD jets fakes a very clean photon signature through a pion or lambda decay. This contamination has been determined to have a momentum balance discrepancy compared to the photon plus jets signal at the 1\% level, and so an additional 0.3\% uncertainty is taken on the track jet momentum. Finally, after applying the procedures described in section~\ref{sec:corrections} to minimize (in the case of \bbbar) or correct (in the case of \ttbar) for jet overlap and underlying event effects, the mass is reevaluated using a cone size of $\Delta R = 0.7$\ instead of 0.4 for the track jets. The resulting shift in the mass results is taken as an additional systematic to represent out-of-cone and jet overlap uncertainties. 

Another track jet energy uncertainty arises in connection with the simulation of the \Bot-jets. If the EvtGen decay tables do not produce the correct distributions of charged particles then this will artificially bias any measurements of the tracking energy of the \Bot-jets. The DELPHI Collaboration has measured the charged decay multiplicity of \Bot-hadrons \cite{bib:DELPHI_b_charge_ref}, excluding the decay products of long lived light-flavor particles and of excited \Bot-hadron to ground state \Bot-hadron transitions, and determined an average of $4.97 \pm 0.07$. We evaluate this number at generator level in our samples using the same exclusions and arrive at a mean result of 5.05. This discrepancy is very slightly larger than the reported error at DELPHI, and it cannot be explained by uncertainties in the production fractions of different \Bot-hadron types or on the semileptonic decay rate. Under the assumption that excess tracks will be distributed randomly in the \Bot-hadron rest frame, this discrepancy should directly translate into an equivalent discrepancy on the measured energy of the component of the track jet originating from the \Bot-hadron decay. This leads to an additional 1.1\% uncertainty on the measured track jet energies. 

In addition to jet energy effects, other uncertainties are considered in relation to the modeling of the physics of the~\bbbar\ sample. It is important to minimize and understand any charm contamination in this sample. Our studies demonstrate that the muon jets in the sample are about 95\% likely to be $b$-jets, and 5\% likely to be charm jets. But they also suggest that the simulation slightly underestimates the number of charm jets. This requires a minor correction of the decay length measured in this calibration sample, and the charm fraction is then fluctuated within its fitted one sigma uncertainties. The small resulting mass shifts are then taken as a systematic uncertainty. Another small uncertainty arises from the event selection on the muons in our leptonic \bbbar\ sample. If the simulation does not properly model the measurement of the muon momentum then higher or lower energy muons (corresponding to higher or lower decay length vertices) will pass selection. While our $Z$ peak fits above suggest a very accurate modeling of isolated muons, we conservatively take a 1\% uncertainty on the muon momentum scale, and evaluate the mass shift that results from the new set of events passing selection. Finally, it is important to understand the uncertainties in \textsc{pythia}'s modeling of the \Bot-quark fragmentation. In \textsc{pythia}, the energy carried by the hadron after the fragmentation process is modeled with the Bowler function. The D0 Collaboration has studied LEP and SLD data and determined the \textsc{pythia} tune required to reproduce their results \cite{bib:D0_lep_sld_frag_ref} . Samples of \ttbar\ events were generated according to each of these tunes, and the resulting \Bot-hadron energies were found to be about 2\% higher than under the default \textsc{pythia} tune. As expected, this results in a proportionally larger mean decay length of our signal \Bot-hadrons. However, since this effect also occurs in the \bbbar\ samples, the effects almost exactly cancel one another out, illustrating the motivation for our calibration procedure. The fragmentation fluctuations do, however, produce the following minor fluctuations which are not canceled out. When events are reweighted to the alternate fragmentation distributions, it causes small alterations to the measured track jet energies and raises the muon energy distribution slightly. Accounting for all of these effects, the larger of the mass shifts between the default \textsc{pythia} sample, and the results after reweighting to the SLD or LEP results are taken as a fragmentation systematic uncertainty . 

A summary of calibration systematic uncertainties for the decay length measurement is shown in Table~\ref{SF_table}.

\begin{table}[th]
  \begin{center}
  \caption{Calibration based top-quark mass uncertainties for the decay length measurement.}\label{SF_table}
  \begin{ruledtabular}
  \begin{tabular}{lccc}
  Systematic [$\textnormal{GeV}/c^2$]& Lxy & Lepton $p_T$ & Simultaneous \\
  \hline
  Photon Plus Jet Stats & 0.7 & 0 & 0.3\\
  Photon $p_T$ & 1.4 & 0 & 0.6 \\
  Photon Background & 0.4 & 0 & 0.2\\
  Track Jet Cone Size & 0.8 & 0 & 0.3\\
  $N_{trk}$ from \Bot-hadrons & 1.6 & 0 & 0.7\\
  \ccbar\ Background & 0.2 & 0 & 0.1\\
  Semilep Muon Pt & 0.3 & 0 & 0.1 \\
  Fragmentation & 0.6 & 0 & 0.3\\
  \hline
  Total Calibration & 2.5 & 0 & 1.1 \\
  \end{tabular}
  \end{ruledtabular}
  \end{center}
\end{table}

\subsection{Multiple interactions uncertainty}

There are two respects in which other interactions during a beam crossing may result in a systematic bias. These extra interactions are simulated as overlaid minimum bias events, however this modeling may not be accurate, resulting in biased jet energies. These effects are described in Section~\ref{sec:cal_JES_uncert} for calorimeter-based jet measurements, and in Section~\ref{sec:Lxy_uncertainties} for the tracking based jet measurements. 

However another effect must be taken into account. The simulation is tuned to an older dataset corresponding to an integrated luminosity of 1.2 \ifb. The newer 0.7 \ifb\ of data included in the measurement were collected at higher instantaneous luminosities with more interactions per bunch crossing than the earlier data. To study this effect we generated a high luminosity \ttbar\ sample with top-quark mass $175\ \textnormal{GeV}/c^2$. We then segregated the events according to the number of collision vertices that are reconstructed in the event (which has been shown to be approximately proportional to the luminosity \cite{bib:newer_CDF_JES_note}). While there is no statistically significant trend over number of vertices for the Lxy measurement, there is a significant dependence for the lepton transverse momentum measurement as shown in Figure~\ref{fig:pileup_fig}. For electrons, a small part of this trend is due to particles from other collisions falling into the electron cluster, however the primary cause of this effect is the isolation requirement for the leptons. Since we require the total calorimeter energy found around the lepton to be less than 10\% of the lepton momentum, low momentum leptons are more likely to fail selection in high luminosity events, as illustrated in Figure~\ref{fig:pileup_fig}. 

Like the simulation, we segregate the data based upon the number of reconstructed collision vertices. As expected, the data have about 15\% more reconstructed collision vertices per event than the standard \ttbar\ samples. The high luminosity simulated events are reweighted to reproduce the distribution of the number of reconstructed vertices in both the standard \ttbar\ samples and the data in turn. These two reweighted results are equivalent to each other within statistics for the Lxy measurement, however the lepton transverse momentum is significantly higher under the luminosity profile of the data. There are insufficient statistics at very high luminosities to reliably correct for this effect in all of the signal \ttbar\ samples. Instead we take the differences between the associated top-quark mass results using the luminosity profiles of the data and the simulation for the high luminosity sample as a systematic uncertainty for each measurement.  We emphasize that this uncertainty is due to the simple logistics of the luminosity profile that was used in the simulation, and is not due to any irreducible physics effect. For the decay length measurement the statistical uncertainty due to the number of generated \ttbar\ events is larger than the observed systematic shift, and so this statistical uncertainty is taken as our systematic instead.

\begin{figure*}
\centerline{
  \subfigure[] {\includegraphics[height=2.5in]{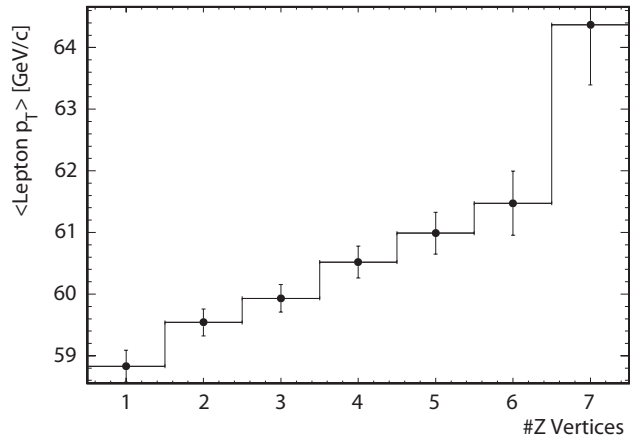}}
}
\caption{Effects of luminosity on the mean lepton transverse momentum. These results are evaluated at generator level and plotted against the number of reconstructed collision vertices. The higher the luminosity is (as measured by the number of vertices), the smaller is the number of low energy leptons that pass the isolation requirement. \label{fig:pileup_fig}}
\end{figure*}

\subsection{Jet energy uncertainties \label{sec:cal_JES_uncert}}

Our jet energy uncertainties can be broken down into two categories: those arising from the tracking energy measurements which impact the Lxy calibration, and those from the calorimeter measurements that are common to all our analyses. None of the uncertainties represent an uncertainty on the determination of a ``true" jet energy. Rather, they represent uncertainties in the modeling of jet energy measurements in simulation. In this section we discuss our evaluation of these uncertainties and explain why they have minimal correlation with the calorimeter-based uncertainties that are claimed by other top-quark mass analyses.

The dominant jet energy uncertainties in this analysis arise from the track jets. They are listed in Table~\ref{SF_table} as Photon Plus Jet Stats, Photon $p_T$, Photon Background, Track Jet Cone Size, and $N_{trk}$ from \Bot-hadrons. Of these, the Photon $p_T$ (energy bias in the calibration photons) and $N_{trk}$ from \Bot-hadrons (EvtGen decay multiplicity mismodeling) categories are the largest contributions to the jet energy uncertainties. The fourth largest uncertainty is due to the limited statistics in the photon plus jets data and will not present any difficulty in future high statistics analyses. 

The third largest uncertainty is due to the size of the cone used to construct our track-based jets. This uncertainty may have components from a wide variety of physical effects, but minimal correlation to the jet energy scale uncertainties of other analyses. The most significant component comes from an uncertainty in the overlap of particles from other jets falling into the jet cone, for which no corresponding uncertainty is claimed for calorimeter jets. The only systematic components for which there are any correlations to calorimeter-based uncertainties are the much smaller underlying event, multiple interaction uncertainties, and out-of-cone uncertainties. Most of the multiple interaction contributions are vetoed by the z-vertex matching requirement. As for the out-of-cone uncertainty, it should be minimal due to the application of the photon calibration procedure. It will only contribute to the extent in which the simulation models out-of-cone effects in \Bot-jets with a different level of accuracy compared to the light flavor jets on which the calibration is performed. To summarize, of all our track jet energy uncertainties, only a small part of the $0.8\ \textnormal{GeV}/c^2$ ($0.3\ \textnormal{GeV}/c^2$) systematic uncertainty on the Lxy (combined) measurements that is due to the altered cone size could have any correlation to the calorimeter-based jet energy scale uncertainties we are claiming, or those of other analyses. 


The calorimeter-based uncertainties are split into six categories, which are assumed to be independent of one another and are described in \cite{bib:newer_CDF_JES_note}.  Since these are the same categories into which the jet energy scale corrections are split, these uncertainties are sometimes called jet energy scale uncertainties. As for the tracking based uncertainties, their impact on the decay length measurement arises based on which jets pass our event selection thresholds. These low energy jets which pass in and out of our sample as the jet energies are varied within uncertainties tend to have a small decay length, and therefore bias the average decay length of our sample. Unlike track jets, however, for calorimeter jets this effect is present in both our \bbbar\ calibration and our main analysis samples, and it largely cancels in the final mass determination.  Such cancellations are what motivated the choice of the Lxy calibration procedure. To evaluate these uncertainties, we fluctuate the calorimeter energies of the jets within these six categories and reevaluate the missing energy of the event, keeping track of which jets and events pass selection. We take the resulting mass shifts as calorimeter-based systematic uncertainties.

One concern that arose at this point is that we may have over-optimized our procedures to create the fortuitous systematic cancellations described above. This concern would specifically pertain to our out-of-cone jet energy uncertainty. The out-of-cone uncertainties are deliberately fixed to be more conservative than the worst case scenario disagreements between \textsc{pythia}-data and \textsc{herwig}-data out-of-cone comparisons as explained in \cite{bib:newer_CDF_JES_note}. But if the out-of-cone disagreement between simulation and data is different for the jets in our \bbbar\ sample than for the jets in our \ttbar\ sample, then by chance the cancellation may lead to an artificially small systematic result. 


To investigate this possibility, we must understand differences between jets in the samples. If jets near the selection threshold were to have identical properties for the \bbbar\ and \ttbar\ samples, then disagreements between data and simulation would be identical for the samples and the resulting systematic cancellation resulting from assuming identical out-of-cone uncertainties would be appropriate. Fortunately, the differences are small. For our purposes, the only relevant differences between the jets in our samples are that the \bbbar\ jets used in our decay length calibration are required to contain muons, and that some of the tagged jets in our backgrounds are light flavor or charm. In all other respects the simulated jets we use near our selection threshold are similar. There are two systematic cross-checks that we run to address these concerns. The results will be shown at the end of this section. 

As explained in \cite{bib:newer_CDF_JES_note}, the out-of-cone jet energy uncertainties are parameterized based upon the calorimeter energy measurement. Since lower energy jets are broader, a lower jet energy corresponds to a larger out-of-cone uncertainty. Since the muon's energy is mostly lost for jets in the \bbbar\ sample, it can be argued that we are overestimating the out-of-cone uncertainty for these jets. To check the impact this would have, we add the muon's energy back in and repeat our systematics evaluation. As a second check, we consider the possibility that the mismodeling of out-of-cone effects by the simulation could be different for heavy and light flavor jets. We check the shifts of the uncertainties which occur when we fluctuate the size of our out-of-cone uncertainties for charm and light flavor jets relative to \Bot-jets within conservative constraints determined by jet shape studies. The fluctuations from these two cross-checks are taken as residual out-of-cone systematic uncertainties. We take our full out-of-cone systematic as the quadrature sum of direct and residual out-of-cone uncertainties. The results are summarized in Table~\ref{cal_jes_table}. The systematic uncertainties from all effects that have been considered are shown in Table~\ref{final_syst_table}.

\begin{table}[th]
  \begin{center}
  \caption{Calorimeter Based Jet Energy Uncertainties. The residual uncertainties result from possible inaccuracies in the cancellation that occurs for our out-of-cone uncertainty.}\label{cal_jes_table}
  \begin{ruledtabular}
  \begin{tabular}{lccc}
  Systematic [$\textnormal{GeV}/c^2$]& Lxy & Lepton $p_T$ & Simultaneous\\
  \hline
  Eta Dependent & 0.06 & -0.08 & -0.02 \\
  Multiple Interactions & 0.17 & -0.01 & 0.07 \\
  Calorimeter Response & -0.14 & -0.07 & -0.09 \\
  Underlying Event & 0.09 & -0.06 & 0.01 \\
  Splash Out & 0.15 & -0.10 & 0.02 \\
  Base Out of Cone & 0.18 & -0.28 & -0.06 \\
  \hline
  \multicolumn{4}{c}{Out of Cone Residual Uncertainties} \\
  \hline
  \bbbar\  Semileptonic &  0.24 & NA & 0.24 \\
  \W\ plus charm/LF &  0.14 & 0.22 & 0.30\\

  Final Out of Cone & 0.33 & 0.36 & 0.24 \\
  \hline
  Total Calorimeter JES & 0.44 & 0.39 & 0.32\\
  \end{tabular}
  \end{ruledtabular}
  \end{center}
\end{table}

\begin{table}[th]
  \begin{center}
  \caption{Final Systematic Uncertainties. The Lxy Calibration systematic is the quadrature sum of the systematics summarized in Table~\ref{SF_table}. The Calorimeter JES systematic is the quadrature sum of the systematics summarized in Table~\ref{cal_jes_table}.}\label{final_syst_table}
  \begin{ruledtabular}
  \begin{tabular}{lccc}
  Systematic [$\textnormal{GeV}/c^2$]& Lxy & Lepton $p_T$ & Simultaneous\\
  \hline
  Background Shape & 1.0 & 2.3 & 1.7 \\
  QCD Radiation & 0.5 & 1.2 & 0.7 \\
  PDF & 0.3 & 0.6 & 0.5\\
  Generator & 0.7 & 0.9 & 0.3 \\
  Lepton $p_T$ Scale & 0 & 2.3 & 1.2 \\
  Lxy Calibration & 2.5 & 0 & 1.1 \\
  Multiple Interactions & 0.2 & 1.2 & 0.7 \\
  Calorimeter JES & 0.4 & 0.4 & 0.3\\
  \hline
  Systematics Total &  2.9 & 3.8 & 2.6 \\
  \end{tabular}
  \end{ruledtabular}
  \end{center}
\end{table}
 
\section{Conclusion}

Using a data sample corresponding to an integrated luminosity of \lumi, we measure a top-quark mass of $m_t = $\ \lxyRes\ using the mean transverse decay length of \Bot-jets, $m_t = $\ \lepPtRes\ using the mean transverse momentum of the leptons from \W-boson decays, and $m_t = $\ \combRes\ using both variables simultaneously. To date, these results represent the most precise measurement of the top-quark mass using an algorithm that has minimal dependence on calorimeter-based jet energy uncertainties. Because we have improved the systematic uncertainty on this measurement by a factor of 3.3 compared to the previous measurement~\cite{bib:lxy_first_measurement}, the precision remains limited by statistics and therefore could improve significantly by the end of Run II. In contrast, if a measurement of this type were performed at the LHC, the systematic uncertainties would be the true limitation as the statistical uncertainties would be negligible.

While a prediction of the systematic uncertainties in such a measurement at the LHC is beyond the scope of this paper, we can predict some of the improvements to the systematics that can be made at the Tevatron. If the simulated events were regenerated using the proper luminosity profile, the multiple interactions uncertainty would become negligibly small. Preliminary studies \cite{bib:athens_latest} have also suggested a method for calibrating the lepton momentum which should improve our lepton $p_T$\ scale systematic by more than a factor of two. If both of these analysis improvements were implemented, the systematic uncertainties for these measurements at CDF would be expected to drop to \lepPtSystLim\ for the lepton transverse momentum measurement, and \combSystLim\ using both variables simultaneously. It appears that an even smaller systematic uncertainty may be attainable by performing the lepton $p_T$\ analysis in the dilepton channel \cite{bib:athens_latest}.


We thank the Fermilab staff and the technical staffs of the participating institutions for their vital contributions. This work was supported by the U.S. Department of Energy and National Science Foundation; the Italian Istituto Nazionale di Fisica Nucleare; the Ministry of Education, Culture, Sports, Science and Technology of Japan; the Natural Sciences and Engineering Research Council of Canada; the National Science Council of the Republic of China; the Swiss National Science Foundation; the A.P. Sloan Foundation; the Bundesministerium f\"ur Bildung und Forschung, Germany; the Korean Science and Engineering Foundation and the Korean Research Foundation; the Science and Technology Facilities Council and the Royal Society, UK; the Institut National de Physique Nucleaire et Physique des Particules/CNRS; the Russian Foundation for Basic Research; the Ministerio de Ciencia e Innovaci\'{o}n, and Programa Consolider-Ingenio 2010, Spain; the Slovak R\&D Agency; and the Academy of Finland. 
\bibliography{prd_v3p0}

\begin{thebibliography}{32}
\bibitem[{bib({\natexlab{d}})}]{bib:CDF_coords}
\bibinfo{note}{The CDF coordinate system is right handed, with z-axis tangent
  to the Tevatron ring and pointing in the direction of the proton beam. The x
  and y coordinates are defined pointing outward and upward from the Tevatron
  ring, respectively. The azimuthal angle $\phi$\ is measured relative to the x
  axis in the transverse plane. The pseudorapidity $\eta$\ is measured from the
  proton direction and is typically expressed as $\eta = -\ln(\tan
  \frac{\theta}{2})$, where $\theta$\ is the polar angle with respect to the
  z-axis. Transverse momentum and energy are defined as $p_T = p \sin(\theta)$
  and $E_T = E \sin(\theta)$, respectively.}
\bibitem[{bib({\natexlab{h}})}]{bib:mean_exp}
\bibinfo{note}{The mean of our Lxy or lepton $p_T$\ distributions can be
  expressed as $<s>f_{s} + <b>f_b$\ where $<s>$\ and $<b>$\ denote the means of
  the signal and background distributions, and $f_s$\ and $f_b$\ represent the
  fractions of events that come from these distributions. Since $<b>$\ does not
  depend on the top quark mass while $<s>$\ depends roughly linearly on the
  top-quark mass, the statistical resolution of the measurement will depend
  linearly on $f_s$.}
\bibitem[{bib({\natexlab{j}})}]{bib:MET_explic}
\bibinfo{note}{More explicitly, the missing transverse energy is defined as
  $|-\sum_i E^i_T \vec{n}_i|$, where the sum is over calorimeter towers with
  transverse energy deposits, $E^i_T$, and $\vec{n}_i$\ is the unit vector from
  the collision vertex to the tower in the plane transverse to the beam
  direction. Further corrections are applied to account for muon tracks and to
  account for calorimeter non-uniformity and non-linearities.}
\end{thebibliography}

\end{document}
%
%